\PassOptionsToPackage{table}{xcolor} %
\documentclass[%
    sigconf,
    natbib=false, %
]{acmart} 

\newbool{isextendedversion}
\setbool{isextendedversion}{true}
\ifbool{isextendedversion}{\newcommand{\extendedversion}[2]{#1}}{\newcommand{\extendedversion}[2]{#2}}

\AtBeginDocument{}

\usepackage[T1]{fontenc} %

\usepackage{hyperref} %

\usepackage{pdflscape}
\usepackage{makecell}
\usepackage{booktabs}
\usepackage[para, flushleft]{threeparttable}
\usepackage{tabularx}
\usepackage{colortbl}
\newcolumntype{L}[1]{>{\raggedright\arraybackslash}p{#1}} %
\newcolumntype{C}[1]{>{\centering\arraybackslash}p{#1}} %
\newcolumntype{R}[1]{>{\raggedleft\arraybackslash}p{#1}} %

\usepackage{float} %
\usepackage{subcaption}

\usepackage{xargs} %
\usepackage{xstring} %

\usepackage{csquotes} %
\DeclareQuoteStyle[american]{english} %
        {\itshape\textquotedblleft}
        [\textquotedblleft]
        {\textquotedblright}
        [0.05em]
        {\textquoteleft}
        {\textquoteright}

\usepackage{tikz}
\usetikzlibrary{shapes,arrows, arrows.meta,matrix, positioning, fit, calc, decorations.pathreplacing}

\newbool{displaycomments}
\setbool{displaycomments}{true}

\ifbool{displaycomments}{%
    \definecolor{springgreen}{rgb}{0.0, 1.0, 0.5}
    \definecolor{dandelion}{HTML}{f0e130}
    \newcommand\COMMENTX[2]{{\bf({#1}: {#2})}}
    \newcommand\sascha[1]{\COMMENTX{sascha}{{{\color{orange}#1}}}}
    \newcommand\chris[1]{\COMMENTX{chris}{{{\color{cyan}#1}}}}
    \newcommand\yas[1]{\COMMENTX{yas}{{{\color{pink}#1}}}}
    \newcommand\jan[1]{\COMMENTX{jan}{{{\color{red}#1}}}}
    \newcommand\marco[1]{\COMMENTX{marco}{{{\color{blue}#1}}}}
    \newcommand\angela[1]{\COMMENTX{angela}{{{\color{green}#1}}}}
    \newcommand\supervisor[1]{\COMMENTX{for supervisor}{{{\color{purple}#1}}}}
    
    \newcommand\todo[1]{\COMMENTX{todo}{{{\color{red}#1}}}}
    \newcommand\draft[1]{{{\color{gray}#1}}}
    \newcommand\red[1]{{{\color{red}#1}}}
}{%
    \newcommand\sascha[1]{}
    \newcommand\chris[1]{}
    \newcommand\yas[1]{}
    \newcommand\jan[1]{}
    \newcommand\marco[1]{}
    \newcommand\angela[1]{}
    \newcommand\supervisor[1]{}
    \newcommand\todo[1]{}
    \newcommand\draft[1]{}
    \newcommand\red[1]{}
}

\newcommand{\boldparagraph}[1]{\paragraph{#1}}
\makeatletter
\renewcommand\boldparagraph{\@startsection{paragraph}{4}{0\parindent}%
    {0.6ex plus 0.6ex minus 0.2ex}%
    {0ex}%
    {\normalfont\normalsize\bfseries\maybe@addperiod}%
}
\newcommand{\maybe@addperiod}[1]{%
    \let\@period\@empty%
    \def\@IEEEsectpunct{}%
    #1\@addpunct{.}\enspace%
}
\makeatother

\usepackage[usetwoside=true]{mdframed}
\usepackage{tcolorbox}
\tcbuselibrary{breakable}
\tcbuselibrary{skins}
\usepackage{enumitem}
\definecolor{darkgray}{gray}{0.3}
\tcbset{}
\newtcolorbox{summaryBox}[2][]
{
    enhanced,
    breakable,
    frame hidden,
    borderline west = {3pt}{0pt}{lightgray},
    colback         = white,
    size            = fbox,
    left            = 0.3em,
    enlarge top by  = 0.2em,%
    coltitle        = black,
    title           = {\color{darkgray} \textbf{#2.} },
    attach title to upper,
    fontupper=\small, 
    #1,
}

\newlength\bubblesize
\newcommand{\ie}{i.e.,}
\newcommand{\eg}{e.g.,}

\newcommand\definevar[2]{%
  \expandafter\newcommand\csname var#1var\endcsname{#2}%
}
\newcommand{\var}[1]{\ifcsname var#1var\endcsname%
        \csname var#1var\endcsname%
    \else\PackageWarning{Var}{`#1' does not exist.}%
        \red{TODO}%
    \fi%
}

\usepackage[
    backend=biber,
    isbn=false,
    sortcites=true, %
    maxbibnames=99, %
    datamodel=acmdatamodel,
    style=acmnumeric,
    ]{biblatex}

\DeclareFieldFormat{titlecase}{#1}

\DeclareCiteCommand{\citewithyear}
  {\usebibmacro{prenote}%
   \bibopenbracket}%
  {\usebibmacro{citeindex}%
   \usebibmacro{cite}
   (\usebibmacro{year})}
  {\multicitedelim}
  {\usebibmacro{postnote}\bibclosebracket}

\definevar{CorpusDocumentCountTotal}{406}
\definevar{DistinctAdviceCount}{1,208}
\definevar{ScientificSourcesCount}{31}
\definevar{ScientificSourcesPercent}{7.6}
\definevar{VideoSourcesCount}{4}
\definevar{VideoSourcesPercent}{1.0}
\definevar{UnavailableSourcesCount}{1}
\definevar{UnavailableSourcesPercent}{0.2}
\definevar{UnrelatedSourcesCount}{29}
\definevar{UnrelatedSourcesPercent}{7.1}
\definevar{DuplicateSourcesCount}{1}
\definevar{DuplicateSourcesPercent}{0.2}
\definevar{NoAdviceSourcesCount}{119}
\definevar{NoAdviceSourcesPercent}{29.3}
\definevar{CodeSnippetsAllDocumentCount}{102}
\definevar{CodeSnippetsAllDocumentPercent}{25.1}
\definevar{LibRecommendationAllDocumentCount}{67}
\definevar{LibRecommendationAllDocumentPercent}{16.5}
\definevar{AbstractAdviceAllDocumentCount}{160}
\definevar{AbstractAdviceAllDocumentPercent}{39.4}
\definevar{CorpusDocumentCountTotal-ScientificSourcesCount-VideoSourcesCount-UnavailableSourcesCount-NoAdviceSources-UnrelatedSourcesCount-DuplicateSourcesCount}{221}
\definevar{TotalRelevantCodesInDocuments}{655}
\definevar{UniqueDocumentsWithRelevantCodes}{124}
\definevar{DistinctAdviceCountRelevant}{272}
\definevar{DistinctAdviceCountRelevantPercent}{22.5}
\definevar{CodeSnippetsOnlyRelevantDocumentCount}{29}
\definevar{CodeSnippetsOnlyRelevantDocumentPercent}{23.4}
\definevar{LibRecommendationOnlyRelevantDocumentCount}{31}
\definevar{LibRecommendationOnlyRelevantDocumentPercent}{25.0}
\definevar{AbstractAdviceOnlyRelevantDocumentCount}{83}
\definevar{AbstractAdviceOnlyRelevantDocumentPercent}{66.9}
\definevar{UniqueDocumentsWithCodesIncludingAuth}{136}
\definevar{UnrelevantAdviceCount}{936}
\definevar{UnrelevantNotSecurityAdviceCount}{257}
\definevar{UnrelevantNotUsabilityAdviceCount}{669}
\definevar{UnrelevantNotAuthAdviceCount}{73}
\definevar{UnrelevantNotAuthDistinctAdviceCount}{50}
\definevar{UnrelevantNotAuthDocumentCount}{41}
\definevar{CorpusDocumentCountTotalPDFPageCount}{4,058}
\definevar{documents.type.blog/news}{219}
\definevar{documents.type.docu/wiki}{79}
\definevar{documents.type.projectwebsite/repo}{41}
\definevar{documents.type.video/audio}{3}
\definevar{documents.type.course/education}{4}
\definevar{documents.type.paper/scientific}{35}
\definevar{documents.type.forum/q&a}{15}
\definevar{documents.type.notfound}{4}
\definevar{documents.type.nothing}{6}
\definevar{documents.type.blog/news.perc}{53.9\%}
\definevar{documents.type.docu/wiki.perc}{19.5\%}
\definevar{documents.type.projectwebsite/repo.perc}{10.1\%}
\definevar{documents.type.video/audio.perc}{0.7\%}
\definevar{documents.type.course/education.perc}{1.0\%}
\definevar{documents.type.paper/scientific.perc}{8.6\%}
\definevar{documents.type.forum/q&a.perc}{3.7\%}
\definevar{documents.type.notfound.perc}{1.0\%}
\definevar{documents.type.nothing.perc}{1.5\%}
\definevar{advice.contradicting.total}{49}
\definevar{advice.contradicting.unique}{49}
\definevar{advice.contradicting.groups}{15}
\definevar{advice.contradicting.documents}{60}
\definevar{advice.contradicting.documents.percent}{48.4}
\definevar{advice.contradicting.conflicting_advice_in_same_document}{2}
\definevar{advice.contradicting.group.c2.advice}{4}
\definevar{advice.contradicting.group.c2.documents}{10}
\definevar{advice.contradicting.group.c13.advice}{6}
\definevar{advice.contradicting.group.c13.documents}{16}
\definevar{advice.contradicting.group.c7.advice}{6}
\definevar{advice.contradicting.group.c7.documents}{9}
\definevar{advice.contradicting.group.c10.advice}{4}
\definevar{advice.contradicting.group.c10.documents}{4}
\definevar{advice.contradicting.group.c9.advice}{2}
\definevar{advice.contradicting.group.c9.documents}{6}
\definevar{advice.contradicting.group.c8.advice}{4}
\definevar{advice.contradicting.group.c8.documents}{2}
\definevar{advice.contradicting.group.c3.advice}{5}
\definevar{advice.contradicting.group.c3.documents}{7}
\definevar{advice.contradicting.group.c1.advice}{2}
\definevar{advice.contradicting.group.c1.documents}{8}
\definevar{advice.contradicting.group.c12.advice}{3}
\definevar{advice.contradicting.group.c12.documents}{5}
\definevar{advice.contradicting.group.c4.advice}{2}
\definevar{advice.contradicting.group.c4.documents}{6}
\definevar{advice.contradicting.group.c5.advice}{2}
\definevar{advice.contradicting.group.c5.documents}{3}
\definevar{advice.contradicting.group.c11.advice}{3}
\definevar{advice.contradicting.group.c11.documents}{7}
\definevar{advice.contradicting.group.c15.advice}{2}
\definevar{advice.contradicting.group.c15.documents}{2}
\definevar{advice.contradicting.group.c6.advice}{2}
\definevar{advice.contradicting.group.c6.documents}{5}
\definevar{advice.contradicting.group.c14.advice}{2}
\definevar{advice.contradicting.group.c14.documents}{6}
\definevar{advice.correctness.undecided}{65}
\definevar{advice.correctness.undecided.perc}{23.9\%}
\definevar{advice.correctness.decided}{207}
\definevar{advice.correctness.correct}{182}
\definevar{advice.correctness.correct.perc}{87.9\%}
\definevar{advice.correctness.incorrect}{25}
\definevar{advice.correctness.incorrect.perc}{12.1\%}
\definevar{advice.correctness.incorrect.affectedDocuments}{24}
\definevar{advice.correctness.incorrect.affectedDocuments.perc}{19.4\%}
\definevar{advice.correctness.contradicting.correct}{21}
\definevar{advice.correctness.contradicting.correct.perc}{42.9\%}
\definevar{advice.correctness.contradicting.incorrect}{14}
\definevar{advice.correctness.contradicting.incorrect.perc}{28.6\%}
\definevar{advice.distribution.perDomain.max}{55}
\definevar{advice.distribution.perDomain.mean}{6.6}
\definevar{advice.distribution.perDomain.median}{3.0}
\definevar{advice.distribution.perUpworker.max}{81}
\definevar{advice.distribution.perUpworker.mean}{26.3}
\definevar{advice.distribution.perUpworker.median}{13.0}
\definevar{advice.distribution.perDocument.max}{35}
\definevar{advice.distribution.perDocument.mean}{4.5}
\definevar{advice.distribution.perDocument.median}{2.0}
\definevar{advice.distribution.frequency.max}{35}
\definevar{advice.distribution.frequency.mean}{2.1}
\definevar{advice.distribution.frequency.median}{1.0}
\definevar{advice.distribution.frequency.onlyOneDocument}{176}
\definevar{advice.distribution.topicPerDocument.max}{9}
\definevar{advice.distribution.topicPerDocument.mean}{3.2}
\definevar{advice.distribution.topicPerDocument.median}{3.0}
\definevar{advice.distribution.topicAuthTypePerDocument.max}{4}
\definevar{advice.distribution.topicAuthTypePerDocument.mean}{1.7}
\definevar{advice.distribution.topicAuthTypePerDocument.median}{1.0}
\definevar{advice.distribution.topicAuthTypePerDocument.onlyOneTopic}{61}
\definevar{advice.questionable}{49}
\definevar{advice.questionable.perc}{18.0\%}
\definevar{advice.questionable.affectedDocuments}{39}
\definevar{advice.questionable.affectedDocuments.perc}{31.5\%}
\definevar{advice.questionable.groups}{14}
\definevar{advice.questionable.q1.advice}{16}
\definevar{advice.questionable.q1.documents}{22}
\definevar{advice.questionable.q4.advice}{2}
\definevar{advice.questionable.q4.documents}{5}
\definevar{advice.questionable.q2.advice}{8}
\definevar{advice.questionable.q2.documents}{10}
\definevar{advice.questionable.q6.advice}{3}
\definevar{advice.questionable.q6.documents}{3}
\definevar{advice.questionable.q3.advice}{3}
\definevar{advice.questionable.q3.documents}{5}
\definevar{advice.questionable.q11.advice}{1}
\definevar{advice.questionable.q11.documents}{1}
\definevar{advice.questionable.q10.advice}{3}
\definevar{advice.questionable.q10.documents}{1}
\definevar{advice.questionable.q7.advice}{2}
\definevar{advice.questionable.q7.documents}{3}
\definevar{advice.questionable.q5.advice}{4}
\definevar{advice.questionable.q5.documents}{4}
\definevar{advice.questionable.q12.advice}{1}
\definevar{advice.questionable.q12.documents}{1}
\definevar{advice.questionable.q14.advice}{1}
\definevar{advice.questionable.q14.documents}{1}
\definevar{advice.questionable.q8.advice}{1}
\definevar{advice.questionable.q8.documents}{3}
\definevar{advice.questionable.q13.advice}{1}
\definevar{advice.questionable.q13.documents}{1}
\definevar{advice.questionable.q9.advice}{3}
\definevar{advice.questionable.q9.documents}{1}
\definevar{advice.questionable.contradicting}{15}
\definevar{advice.questionable.contradicting.perc}{30.6\%}
\definevar{advice.recommendable}{86}
\definevar{advice.recommendable.perc}{31.6\%}
\definevar{advice.recommendable.affectedDocuments}{92}
\definevar{advice.recommendable.affectedDocuments.perc}{74.2\%}
\definevar{advice.potentially-recommendable}{40}
\definevar{advice.potentially-recommendable.perc}{14.7\%}
\definevar{advice.potentially-recommendable.affectedDocuments}{46}
\definevar{advice.potentially-recommendable.affectedDocuments.perc}{37.1\%}
\definevar{advice.categories}{12}
\definevar{advice.category.passwords}{264}
\definevar{advice.category.passwords.documents}{64}
\definevar{advice.category.passwords.documents.perc}{51.6\%}
\definevar{advice.category.passwords.unique}{101}
\definevar{advice.category.passwords.mostcommon}{16}
\definevar{advice.category.passwords.mostcommon.documents}{14}
\definevar{advice.category.passwords.mostcommon.advice}{Use/Enforce a password policy.}
\definevar{advice.category.passwords.2ndcommon}{12}
\definevar{advice.category.passwords.2ndcommon.documents}{11}
\definevar{advice.category.passwords.2ndcommon.advice}{Require strong passwords}
\definevar{advice.category.passwords.3rdcommon}{11}
\definevar{advice.category.passwords.3rdcommon.documents}{10}
\definevar{advice.category.passwords.3rdcommon.advice}{Reject most often used passwords/from dictionaries}
\definevar{advice.category.passwords.4thcommon}{9}
\definevar{advice.category.passwords.4thcommon.documents}{6}
\definevar{advice.category.passwords.4thcommon.advice}{Use password alternatives.}
\definevar{advice.category.passwords.5thcommon}{8}
\definevar{advice.category.passwords.5thcommon.documents}{6}
\definevar{advice.category.passwords.5thcommon.advice}{Use password strength meters.}
\definevar{advice.category.2fa/mfa}{87}
\definevar{advice.category.2fa/mfa.documents}{54}
\definevar{advice.category.2fa/mfa.documents.perc}{43.5\%}
\definevar{advice.category.2fa/mfa.unique}{36}
\definevar{advice.category.2fa/mfa.mostcommon}{41}
\definevar{advice.category.2fa/mfa.mostcommon.documents}{35}
\definevar{advice.category.2fa/mfa.mostcommon.advice}{Offer/Use 2FA/MFA}
\definevar{advice.category.2fa/mfa.2ndcommon}{4}
\definevar{advice.category.2fa/mfa.2ndcommon.documents}{4}
\definevar{advice.category.2fa/mfa.2ndcommon.advice}{Don't use SMS as 2FA}
\definevar{advice.category.2fa/mfa.3rdcommon}{4}
\definevar{advice.category.2fa/mfa.3rdcommon.documents}{4}
\definevar{advice.category.2fa/mfa.3rdcommon.advice}{Use adaptive MFA: request more factors if action has high risk}
\definevar{advice.category.2fa/mfa.4thcommon}{2}
\definevar{advice.category.2fa/mfa.4thcommon.documents}{2}
\definevar{advice.category.2fa/mfa.4thcommon.advice}{Use mobile device push notifications as second factor}
\definevar{advice.category.2fa/mfa.5thcommon}{2}
\definevar{advice.category.2fa/mfa.5thcommon.documents}{2}
\definevar{advice.category.2fa/mfa.5thcommon.advice}{Provide OTPs as 2FA option}
\definevar{advice.category.sso}{50}
\definevar{advice.category.sso.documents}{28}
\definevar{advice.category.sso.documents.perc}{22.6\%}
\definevar{advice.category.sso.unique}{20}
\definevar{advice.category.sso.mostcommon}{16}
\definevar{advice.category.sso.mostcommon.documents}{14}
\definevar{advice.category.sso.mostcommon.advice}{Allow/Use 3rd-party SSO.}
\definevar{advice.category.sso.2ndcommon}{3}
\definevar{advice.category.sso.2ndcommon.documents}{3}
\definevar{advice.category.sso.2ndcommon.advice}{Use OpenID Connect for authentication.}
\definevar{advice.category.sso.3rdcommon}{3}
\definevar{advice.category.sso.3rdcommon.documents}{3}
\definevar{advice.category.sso.3rdcommon.advice}{Use SAML for enterprise applications.}
\definevar{advice.category.sso.4thcommon}{3}
\definevar{advice.category.sso.4thcommon.documents}{3}
\definevar{advice.category.sso.4thcommon.advice}{Implement 3rd-party authentication via OAuth.}
\definevar{advice.category.sso.5thcommon}{3}
\definevar{advice.category.sso.5thcommon.documents}{3}
\definevar{advice.category.sso.5thcommon.advice}{Allow to link/merge SSO accounts of the same person}
\definevar{advice.category.bio}{11}
\definevar{advice.category.bio.documents}{5}
\definevar{advice.category.bio.documents.perc}{4.0\%}
\definevar{advice.category.bio.unique}{8}
\definevar{advice.category.bio.mostcommon}{3}
\definevar{advice.category.bio.mostcommon.documents}{2}
\definevar{advice.category.bio.mostcommon.advice}{Take environmental factors into consideration (biometrics)}
\definevar{advice.category.bio.2ndcommon}{2}
\definevar{advice.category.bio.2ndcommon.documents}{1}
\definevar{advice.category.bio.2ndcommon.advice}{Use fingerprints as userid for medical records}
\definevar{advice.category.bio.3rdcommon}{1}
\definevar{advice.category.bio.3rdcommon.documents}{1}
\definevar{advice.category.bio.3rdcommon.advice}{Protect 2FA with PINs or biometrics}
\definevar{advice.category.bio.4thcommon}{1}
\definevar{advice.category.bio.4thcommon.documents}{1}
\definevar{advice.category.bio.4thcommon.advice}{Chose only biometrics that fit your application}
\definevar{advice.category.bio.5thcommon}{1}
\definevar{advice.category.bio.5thcommon.documents}{1}
\definevar{advice.category.bio.5thcommon.advice}{Prefer fingerprint scanning above facial recognition}
\definevar{advice.category.token}{23}
\definevar{advice.category.token.documents}{12}
\definevar{advice.category.token.documents.perc}{9.7\%}
\definevar{advice.category.token.unique}{16}
\definevar{advice.category.token.mostcommon}{5}
\definevar{advice.category.token.mostcommon.documents}{5}
\definevar{advice.category.token.mostcommon.advice}{Give tokens an expiration}
\definevar{advice.category.token.2ndcommon}{2}
\definevar{advice.category.token.2ndcommon.documents}{2}
\definevar{advice.category.token.2ndcommon.advice}{Support token-based auth.}
\definevar{advice.category.token.3rdcommon}{2}
\definevar{advice.category.token.3rdcommon.documents}{2}
\definevar{advice.category.token.3rdcommon.advice}{Use cookie-based auth for web and mobile.}
\definevar{advice.category.token.4thcommon}{2}
\definevar{advice.category.token.4thcommon.documents}{2}
\definevar{advice.category.token.4thcommon.advice}{Use cookie-based auth for web-only applications.}
\definevar{advice.category.token.5thcommon}{1}
\definevar{advice.category.token.5thcommon.documents}{1}
\definevar{advice.category.token.5thcommon.advice}{Force your client/user to change password immediately}
\definevar{advice.category.other}{26}
\definevar{advice.category.other.documents}{14}
\definevar{advice.category.other.documents.perc}{11.3\%}
\definevar{advice.category.other.unique}{10}
\definevar{advice.category.other.mostcommon}{9}
\definevar{advice.category.other.mostcommon.documents}{6}
\definevar{advice.category.other.mostcommon.advice}{Use password alternatives.}
\definevar{advice.category.other.2ndcommon}{5}
\definevar{advice.category.other.2ndcommon.documents}{5}
\definevar{advice.category.other.2ndcommon.advice}{Use email authentication/one-time login links}
\definevar{advice.category.other.3rdcommon}{2}
\definevar{advice.category.other.3rdcommon.documents}{2}
\definevar{advice.category.other.3rdcommon.advice}{Use TLS Client Authentication, if}
\definevar{advice.category.other.4thcommon}{2}
\definevar{advice.category.other.4thcommon.documents}{2}
\definevar{advice.category.other.4thcommon.advice}{Use TLS Client Authentication if extra security required}
\definevar{advice.category.other.5thcommon}{2}
\definevar{advice.category.other.5thcommon.documents}{2}
\definevar{advice.category.other.5thcommon.advice}{Consider using a security question/memorable words}
\definevar{advice.category.ui}{104}
\definevar{advice.category.ui.documents}{35}
\definevar{advice.category.ui.documents.perc}{28.2\%}
\definevar{advice.category.ui.unique}{50}
\definevar{advice.category.ui.mostcommon}{19}
\definevar{advice.category.ui.mostcommon.documents}{13}
\definevar{advice.category.ui.mostcommon.advice}{Use generic responses for auth errors to not weaken security}
\definevar{advice.category.ui.2ndcommon}{8}
\definevar{advice.category.ui.2ndcommon.documents}{6}
\definevar{advice.category.ui.2ndcommon.advice}{Use password strength meters.}
\definevar{advice.category.ui.3rdcommon}{9}
\definevar{advice.category.ui.3rdcommon.documents}{5}
\definevar{advice.category.ui.3rdcommon.advice}{Allow users to toggle password visibility when typing it}
\definevar{advice.category.ui.4thcommon}{5}
\definevar{advice.category.ui.4thcommon.documents}{4}
\definevar{advice.category.ui.4thcommon.advice}{Keep users signed in.}
\definevar{advice.category.ui.5thcommon}{4}
\definevar{advice.category.ui.5thcommon.documents}{2}
\definevar{advice.category.ui.5thcommon.advice}{Use "login" and "sign up"; not "sign in" and "sign up".}
\definevar{advice.category.sessaccmngmnt}{225}
\definevar{advice.category.sessaccmngmnt.documents}{63}
\definevar{advice.category.sessaccmngmnt.documents.perc}{50.8\%}
\definevar{advice.category.sessaccmngmnt.unique}{103}
\definevar{advice.category.sessaccmngmnt.mostcommon}{13}
\definevar{advice.category.sessaccmngmnt.mostcommon.documents}{10}
\definevar{advice.category.sessaccmngmnt.mostcommon.advice}{Require re-authentication for sensitive actions (step-up auth).}
\definevar{advice.category.sessaccmngmnt.2ndcommon}{9}
\definevar{advice.category.sessaccmngmnt.2ndcommon.documents}{7}
\definevar{advice.category.sessaccmngmnt.2ndcommon.advice}{Limit the number of login attempts.}
\definevar{advice.category.sessaccmngmnt.3rdcommon}{8}
\definevar{advice.category.sessaccmngmnt.3rdcommon.documents}{7}
\definevar{advice.category.sessaccmngmnt.3rdcommon.advice}{Protect against brute-force attacks by rate limiting}
\definevar{advice.category.sessaccmngmnt.4thcommon}{7}
\definevar{advice.category.sessaccmngmnt.4thcommon.documents}{6}
\definevar{advice.category.sessaccmngmnt.4thcommon.advice}{Provide secure password recovery mechanism}
\definevar{advice.category.sessaccmngmnt.5thcommon}{6}
\definevar{advice.category.sessaccmngmnt.5thcommon.documents}{6}
\definevar{advice.category.sessaccmngmnt.5thcommon.advice}{Enforce users to change passwords regularly}
\definevar{advice.category.usage}{188}
\definevar{advice.category.usage.documents}{83}
\definevar{advice.category.usage.documents.perc}{66.9\%}
\definevar{advice.category.usage.unique}{71}
\definevar{advice.category.usage.mostcommon}{41}
\definevar{advice.category.usage.mostcommon.documents}{35}
\definevar{advice.category.usage.mostcommon.advice}{Offer/Use 2FA/MFA}
\definevar{advice.category.usage.2ndcommon}{16}
\definevar{advice.category.usage.2ndcommon.documents}{14}
\definevar{advice.category.usage.2ndcommon.advice}{Allow/Use 3rd-party SSO.}
\definevar{advice.category.usage.3rdcommon}{13}
\definevar{advice.category.usage.3rdcommon.documents}{10}
\definevar{advice.category.usage.3rdcommon.advice}{Require re-authentication for sensitive actions (step-up auth).}
\definevar{advice.category.usage.4thcommon}{9}
\definevar{advice.category.usage.4thcommon.documents}{6}
\definevar{advice.category.usage.4thcommon.advice}{Use password alternatives.}
\definevar{advice.category.usage.5thcommon}{5}
\definevar{advice.category.usage.5thcommon.documents}{5}
\definevar{advice.category.usage.5thcommon.advice}{Use email authentication/one-time login links}
\definevar{advice.category.notification}{59}
\definevar{advice.category.notification.documents}{27}
\definevar{advice.category.notification.documents.perc}{21.8\%}
\definevar{advice.category.notification.unique}{27}
\definevar{advice.category.notification.mostcommon}{19}
\definevar{advice.category.notification.mostcommon.documents}{13}
\definevar{advice.category.notification.mostcommon.advice}{Use generic responses for auth errors to not weaken security}
\definevar{advice.category.notification.2ndcommon}{9}
\definevar{advice.category.notification.2ndcommon.documents}{4}
\definevar{advice.category.notification.2ndcommon.advice}{Notify users of suspicious login behavior/failed attempts}
\definevar{advice.category.notification.3rdcommon}{2}
\definevar{advice.category.notification.3rdcommon.documents}{2}
\definevar{advice.category.notification.3rdcommon.advice}{Use mobile device push notifications as second factor}
\definevar{advice.category.notification.4thcommon}{2}
\definevar{advice.category.notification.4thcommon.documents}{2}
\definevar{advice.category.notification.4thcommon.advice}{Never send the old password to the user}
\definevar{advice.category.notification.5thcommon}{2}
\definevar{advice.category.notification.5thcommon.documents}{2}
\definevar{advice.category.notification.5thcommon.advice}{Notify users via mail when password was reset.}
\definevar{advice.category.education}{12}
\definevar{advice.category.education.documents}{8}
\definevar{advice.category.education.documents.perc}{6.5\%}
\definevar{advice.category.education.unique}{7}
\definevar{advice.category.education.mostcommon}{5}
\definevar{advice.category.education.mostcommon.documents}{4}
\definevar{advice.category.education.mostcommon.advice}{Educate users to create strong passwords.}
\definevar{advice.category.education.2ndcommon}{2}
\definevar{advice.category.education.2ndcommon.documents}{1}
\definevar{advice.category.education.2ndcommon.advice}{Provide security advice to users.}
\definevar{advice.category.education.3rdcommon}{1}
\definevar{advice.category.education.3rdcommon.documents}{1}
\definevar{advice.category.education.3rdcommon.advice}{Encourage users to register multiple 2FAs for account recovery}
\definevar{advice.category.education.4thcommon}{1}
\definevar{advice.category.education.4thcommon.documents}{1}
\definevar{advice.category.education.4thcommon.advice}{Educate instead of enforcing}
\definevar{advice.category.education.5thcommon}{1}
\definevar{advice.category.education.5thcommon.documents}{1}
\definevar{advice.category.education.5thcommon.advice}{Educate users to use password managers.}
\definevar{advice.category.misc}{18}
\definevar{advice.category.misc.documents}{10}
\definevar{advice.category.misc.documents.perc}{8.1\%}
\definevar{advice.category.misc.unique}{9}
\definevar{advice.category.misc.mostcommon}{3}
\definevar{advice.category.misc.mostcommon.documents}{3}
\definevar{advice.category.misc.mostcommon.advice}{Use CAPTCHAs to make automated attempts more expensive}
\definevar{advice.category.misc.2ndcommon}{2}
\definevar{advice.category.misc.2ndcommon.documents}{2}
\definevar{advice.category.misc.2ndcommon.advice}{Use CAPTCHAs after failed attempts}
\definevar{advice.category.misc.3rdcommon}{2}
\definevar{advice.category.misc.3rdcommon.documents}{2}
\definevar{advice.category.misc.3rdcommon.advice}{Provide captcha after limit is reached}
\definevar{advice.category.misc.4thcommon}{5}
\definevar{advice.category.misc.4thcommon.documents}{1}
\definevar{advice.category.misc.4thcommon.advice}{Provide methods to revoke access for sharing PII}
\definevar{advice.category.misc.5thcommon}{2}
\definevar{advice.category.misc.5thcommon.documents}{1}
\definevar{advice.category.misc.5thcommon.advice}{Involve stakeholders early on for security requirements}

\definevar{participants.Upwork}{18}
\definevar{participants.Pilot}{9}
\definevar{participants.Upwork.text}{18}
\definevar{participants.Upwork.Text}{18}
\definevar{participants.Pilot.text}{nine}
\definevar{participants.Pilot.Text}{Nine}

\definevar{demo.age.Prefernottodisclose}{0}
\definevar{demo.age.count}{18.0}
\definevar{demo.age.mean}{28.4}
\definevar{demo.age.std}{7.9}
\definevar{demo.age.min}{19.0}
\definevar{demo.age.25percentile}{24.0}
\definevar{demo.age.50percentile}{25.5}
\definevar{demo.age.75percentile}{31.5}
\definevar{demo.age.max}{45.0}
\definevar{demo.gender.Prefernottodisclose}{0}
\definevar{demo.gender.Man}{16}
\definevar{demo.gender.Woman}{2}
\definevar{demo.gender.Man.perc}{88.9\%}
\definevar{demo.gender.Woman.perc}{11.1\%}
\definevar{demo.exp.Prefernottodisclose}{0}
\definevar{demo.exp.count}{18.0}
\definevar{demo.exp.mean}{6.1}
\definevar{demo.exp.std}{6.0}
\definevar{demo.exp.min}{1.0}
\definevar{demo.exp.25percentile}{2.0}
\definevar{demo.exp.50percentile}{3.8}
\definevar{demo.exp.75percentile}{6.8}
\definevar{demo.exp.max}{22.0}
\definevar{demo.countries.Prefernottodisclose}{0}
\definevar{demo.countries.UnitedKingdom}{4}
\definevar{demo.countries.UnitedStates}{3}
\definevar{demo.countries.India}{2}
\definevar{demo.countries.Brazil}{2}
\definevar{demo.countries.Canada}{1}
\definevar{demo.countries.Nepal}{1}
\definevar{demo.countries.Romania}{1}
\definevar{demo.countries.Germany}{1}
\definevar{demo.countries.Nigeria}{1}
\definevar{demo.countries.France}{1}
\definevar{demo.countries.Turkey}{1}
\definevar{demo.countries.UnitedKingdom.perc}{22.2\%}
\definevar{demo.countries.UnitedStates.perc}{16.7\%}
\definevar{demo.countries.India.perc}{11.1\%}
\definevar{demo.countries.Brazil.perc}{11.1\%}
\definevar{demo.countries.Canada.perc}{5.6\%}
\definevar{demo.countries.Nepal.perc}{5.6\%}
\definevar{demo.countries.Romania.perc}{5.6\%}
\definevar{demo.countries.Germany.perc}{5.6\%}
\definevar{demo.countries.Nigeria.perc}{5.6\%}
\definevar{demo.countries.France.perc}{5.6\%}
\definevar{demo.countries.Turkey.perc}{5.6\%}
\definevar{demo.countries.distinct}{11}
\definevar{demo.countries.other}{7}
\definevar{demo.countries.other.perc}{38.9\%}
\definevar{demo.employment.Prefernottodisclose}{0}
\definevar{demo.employment.Employedfull-time}{10}
\definevar{demo.employment.Self-employed/Freelancer}{9}
\definevar{demo.employment.Employedpart-time}{2}
\definevar{demo.employment.Student}{2}
\definevar{demo.employment.Prefertoself-describe}{1}
\definevar{demo.employment.Employedfull-time.perc}{55.6\%}
\definevar{demo.employment.Self-employed/Freelancer.perc}{50.0\%}
\definevar{demo.employment.Employedpart-time.perc}{11.1\%}
\definevar{demo.employment.Student.perc}{11.1\%}
\definevar{demo.employment.Prefertoself-describe.perc}{5.6\%}
\definevar{demo.education.Prefernottodisclose}{0}
\definevar{demo.education.Bachelor'sdegree}{7}
\definevar{demo.education.Highschoolorequivalent/Alevelorequivalent}{4}
\definevar{demo.education.Master'sorprofessionaldegree}{3}
\definevar{demo.education.Somecollege(orcurrentlyenrolled),ortwo-yearassociate’sdegree}{2}
\definevar{demo.education.Lessthanhighschool/GCSEorequivalent}{1}
\definevar{demo.education.Somegraduateschool,orcurrentlyenrolledingraduateschool}{1}
\definevar{demo.education.Bachelor'sdegree.perc}{38.9\%}
\definevar{demo.education.Highschoolorequivalent/Alevelorequivalent.perc}{22.2\%}
\definevar{demo.education.Master'sorprofessionaldegree.perc}{16.7\%}
\definevar{demo.education.Somecollege(orcurrentlyenrolled),ortwo-yearassociate’sdegree.perc}{11.1\%}
\definevar{demo.education.Lessthanhighschool/GCSEorequivalent.perc}{5.6\%}
\definevar{demo.education.Somegraduateschool,orcurrentlyenrolledingraduateschool.perc}{5.6\%}
\definevar{demo.webDomainExp.Prefernottodisclose}{0}
\definevar{demo.webDomainExp.Full-StackDevelopment}{15}
\definevar{demo.webDomainExp.FrontendDevelopment}{11}
\definevar{demo.webDomainExp.BackendDevelopment}{9}
\definevar{demo.webDomainExp.WebDesign}{7}
\definevar{demo.webDomainExp.UI/UX}{6}
\definevar{demo.webDomainExp.APIDevelopment}{6}
\definevar{demo.webDomainExp.Testing}{4}
\definevar{demo.webDomainExp.Other(pleasespecifybelow)}{1}
\definevar{demo.webDomainExp.Full-StackDevelopment.perc}{83.3\%}
\definevar{demo.webDomainExp.FrontendDevelopment.perc}{61.1\%}
\definevar{demo.webDomainExp.BackendDevelopment.perc}{50.0\%}
\definevar{demo.webDomainExp.WebDesign.perc}{38.9\%}
\definevar{demo.webDomainExp.UI/UX.perc}{33.3\%}
\definevar{demo.webDomainExp.APIDevelopment.perc}{33.3\%}
\definevar{demo.webDomainExp.Testing.perc}{22.2\%}
\definevar{demo.webDomainExp.Other(pleasespecifybelow).perc}{5.6\%}
\definevar{demo.usability.experience.Prefernottodisclose}{0}
\definevar{demo.usability.experience.Some}{8}
\definevar{demo.usability.experience.Considerable}{7}
\definevar{demo.usability.experience.Little}{2}
\definevar{demo.usability.experience.No}{1}
\definevar{demo.usability.experience.Some.perc}{44.4\%}
\definevar{demo.usability.experience.Considerable.perc}{38.9\%}
\definevar{demo.usability.experience.Little.perc}{11.1\%}
\definevar{demo.usability.experience.No.perc}{5.6\%}
\definevar{demo.security.experience.Prefernottodisclose}{0}
\definevar{demo.security.experience.Some}{12}
\definevar{demo.security.experience.Considerable}{3}
\definevar{demo.security.experience.Little}{2}
\definevar{demo.security.experience.No}{1}
\definevar{demo.security.experience.Some.perc}{66.7\%}
\definevar{demo.security.experience.Considerable.perc}{16.7\%}
\definevar{demo.security.experience.Little.perc}{11.1\%}
\definevar{demo.security.experience.No.perc}{5.6\%}
\definevar{demo.usability.training.studies.Prefernottodisclose}{0}
\definevar{demo.usability.training.studies.No}{13}
\definevar{demo.usability.training.studies.Yes}{5}
\definevar{demo.usability.training.studies.No.perc}{72.2\%}
\definevar{demo.usability.training.studies.Yes.perc}{27.8\%}
\definevar{demo.security.training.studies.Prefernottodisclose}{0}
\definevar{demo.security.training.studies.No}{10}
\definevar{demo.security.training.studies.Yes}{8}
\definevar{demo.security.training.studies.No.perc}{55.6\%}
\definevar{demo.security.training.studies.Yes.perc}{44.4\%}
\definevar{demo.usability.training.job.Prefernottodisclose}{0}
\definevar{demo.usability.training.job.No}{11}
\definevar{demo.usability.training.job.Yes}{7}
\definevar{demo.usability.training.job.No.perc}{61.1\%}
\definevar{demo.usability.training.job.Yes.perc}{38.9\%}
\definevar{demo.security.training.job.Prefernottodisclose}{0}
\definevar{demo.security.training.job.No}{11}
\definevar{demo.security.training.job.Yes}{7}
\definevar{demo.security.training.job.No.perc}{61.1\%}
\definevar{demo.security.training.job.Yes.perc}{38.9\%}
\definevar{demo.age.Prefernottodisclose.text}{zero}
\definevar{demo.age.Prefernottodisclose.Text}{Zero}
\definevar{demo.age.count.text}{18.0}
\definevar{demo.age.count.Text}{18.0}
\definevar{demo.age.mean.text}{28.4}
\definevar{demo.age.mean.Text}{28.4}
\definevar{demo.age.std.text}{seven}
\definevar{demo.age.std.Text}{Seven}
\definevar{demo.age.min.text}{19.0}
\definevar{demo.age.min.Text}{19.0}
\definevar{demo.age.25percentile.text}{24.0}
\definevar{demo.age.25percentile.Text}{24.0}
\definevar{demo.age.50percentile.text}{25.5}
\definevar{demo.age.50percentile.Text}{25.5}
\definevar{demo.age.75percentile.text}{31.5}
\definevar{demo.age.75percentile.Text}{31.5}
\definevar{demo.age.max.text}{45.0}
\definevar{demo.age.max.Text}{45.0}
\definevar{demo.gender.Prefernottodisclose.text}{zero}
\definevar{demo.gender.Prefernottodisclose.Text}{Zero}
\definevar{demo.gender.Man.text}{16}
\definevar{demo.gender.Man.Text}{16}
\definevar{demo.gender.Woman.text}{two}
\definevar{demo.gender.Woman.Text}{Two}
\definevar{demo.gender.Man.perc.text}{zero}
\definevar{demo.gender.Man.perc.Text}{Zero}
\definevar{demo.gender.Woman.perc.text}{zero}
\definevar{demo.gender.Woman.perc.Text}{Zero}
\definevar{demo.exp.Prefernottodisclose.text}{zero}
\definevar{demo.exp.Prefernottodisclose.Text}{Zero}
\definevar{demo.exp.count.text}{18.0}
\definevar{demo.exp.count.Text}{18.0}
\definevar{demo.exp.mean.text}{six}
\definevar{demo.exp.mean.Text}{Six}
\definevar{demo.exp.std.text}{six}
\definevar{demo.exp.std.Text}{Six}
\definevar{demo.exp.min.text}{one}
\definevar{demo.exp.min.Text}{One}
\definevar{demo.exp.25percentile.text}{two}
\definevar{demo.exp.25percentile.Text}{Two}
\definevar{demo.exp.50percentile.text}{three}
\definevar{demo.exp.50percentile.Text}{Three}
\definevar{demo.exp.75percentile.text}{six}
\definevar{demo.exp.75percentile.Text}{Six}
\definevar{demo.exp.max.text}{22.0}
\definevar{demo.exp.max.Text}{22.0}
\definevar{demo.countries.Prefernottodisclose.text}{zero}
\definevar{demo.countries.Prefernottodisclose.Text}{Zero}
\definevar{demo.countries.UnitedKingdom.text}{four}
\definevar{demo.countries.UnitedKingdom.Text}{Four}
\definevar{demo.countries.UnitedStates.text}{three}
\definevar{demo.countries.UnitedStates.Text}{Three}
\definevar{demo.countries.India.text}{two}
\definevar{demo.countries.India.Text}{Two}
\definevar{demo.countries.Brazil.text}{two}
\definevar{demo.countries.Brazil.Text}{Two}
\definevar{demo.countries.Canada.text}{one}
\definevar{demo.countries.Canada.Text}{One}
\definevar{demo.countries.Nepal.text}{one}
\definevar{demo.countries.Nepal.Text}{One}
\definevar{demo.countries.Romania.text}{one}
\definevar{demo.countries.Romania.Text}{One}
\definevar{demo.countries.Germany.text}{one}
\definevar{demo.countries.Germany.Text}{One}
\definevar{demo.countries.Nigeria.text}{one}
\definevar{demo.countries.Nigeria.Text}{One}
\definevar{demo.countries.France.text}{one}
\definevar{demo.countries.France.Text}{One}
\definevar{demo.countries.Turkey.text}{one}
\definevar{demo.countries.Turkey.Text}{One}
\definevar{demo.countries.UnitedKingdom.perc.text}{zero}
\definevar{demo.countries.UnitedKingdom.perc.Text}{Zero}
\definevar{demo.countries.UnitedStates.perc.text}{zero}
\definevar{demo.countries.UnitedStates.perc.Text}{Zero}
\definevar{demo.countries.India.perc.text}{zero}
\definevar{demo.countries.India.perc.Text}{Zero}
\definevar{demo.countries.Brazil.perc.text}{zero}
\definevar{demo.countries.Brazil.perc.Text}{Zero}
\definevar{demo.countries.Canada.perc.text}{zero}
\definevar{demo.countries.Canada.perc.Text}{Zero}
\definevar{demo.countries.Nepal.perc.text}{zero}
\definevar{demo.countries.Nepal.perc.Text}{Zero}
\definevar{demo.countries.Romania.perc.text}{zero}
\definevar{demo.countries.Romania.perc.Text}{Zero}
\definevar{demo.countries.Germany.perc.text}{zero}
\definevar{demo.countries.Germany.perc.Text}{Zero}
\definevar{demo.countries.Nigeria.perc.text}{zero}
\definevar{demo.countries.Nigeria.perc.Text}{Zero}
\definevar{demo.countries.France.perc.text}{zero}
\definevar{demo.countries.France.perc.Text}{Zero}
\definevar{demo.countries.Turkey.perc.text}{zero}
\definevar{demo.countries.Turkey.perc.Text}{Zero}
\definevar{demo.countries.distinct.text}{eleven}
\definevar{demo.countries.distinct.Text}{Eleven}
\definevar{demo.countries.other.text}{seven}
\definevar{demo.countries.other.Text}{Seven}
\definevar{demo.countries.other.perc.text}{zero}
\definevar{demo.countries.other.perc.Text}{Zero}
\definevar{demo.employment.Prefernottodisclose.text}{zero}
\definevar{demo.employment.Prefernottodisclose.Text}{Zero}
\definevar{demo.employment.Employedfull-time.text}{ten}
\definevar{demo.employment.Employedfull-time.Text}{Ten}
\definevar{demo.employment.Self-employed/Freelancer.text}{nine}
\definevar{demo.employment.Self-employed/Freelancer.Text}{Nine}
\definevar{demo.employment.Employedpart-time.text}{two}
\definevar{demo.employment.Employedpart-time.Text}{Two}
\definevar{demo.employment.Student.text}{two}
\definevar{demo.employment.Student.Text}{Two}
\definevar{demo.employment.Prefertoself-describe.text}{one}
\definevar{demo.employment.Prefertoself-describe.Text}{One}
\definevar{demo.employment.Employedfull-time.perc.text}{zero}
\definevar{demo.employment.Employedfull-time.perc.Text}{Zero}
\definevar{demo.employment.Self-employed/Freelancer.perc.text}{zero}
\definevar{demo.employment.Self-employed/Freelancer.perc.Text}{Zero}
\definevar{demo.employment.Employedpart-time.perc.text}{zero}
\definevar{demo.employment.Employedpart-time.perc.Text}{Zero}
\definevar{demo.employment.Student.perc.text}{zero}
\definevar{demo.employment.Student.perc.Text}{Zero}
\definevar{demo.employment.Prefertoself-describe.perc.text}{zero}
\definevar{demo.employment.Prefertoself-describe.perc.Text}{Zero}
\definevar{demo.education.Prefernottodisclose.text}{zero}
\definevar{demo.education.Prefernottodisclose.Text}{Zero}
\definevar{demo.education.Bachelor'sdegree.text}{seven}
\definevar{demo.education.Bachelor'sdegree.Text}{Seven}
\definevar{demo.education.Highschoolorequivalent/Alevelorequivalent.text}{four}
\definevar{demo.education.Highschoolorequivalent/Alevelorequivalent.Text}{Four}
\definevar{demo.education.Master'sorprofessionaldegree.text}{three}
\definevar{demo.education.Master'sorprofessionaldegree.Text}{Three}
\definevar{demo.education.Somecollege(orcurrentlyenrolled),ortwo-yearassociate’sdegree.text}{two}
\definevar{demo.education.Somecollege(orcurrentlyenrolled),ortwo-yearassociate’sdegree.Text}{Two}
\definevar{demo.education.Lessthanhighschool/GCSEorequivalent.text}{one}
\definevar{demo.education.Lessthanhighschool/GCSEorequivalent.Text}{One}
\definevar{demo.education.Somegraduateschool,orcurrentlyenrolledingraduateschool.text}{one}
\definevar{demo.education.Somegraduateschool,orcurrentlyenrolledingraduateschool.Text}{One}
\definevar{demo.education.Bachelor'sdegree.perc.text}{zero}
\definevar{demo.education.Bachelor'sdegree.perc.Text}{Zero}
\definevar{demo.education.Highschoolorequivalent/Alevelorequivalent.perc.text}{zero}
\definevar{demo.education.Highschoolorequivalent/Alevelorequivalent.perc.Text}{Zero}
\definevar{demo.education.Master'sorprofessionaldegree.perc.text}{zero}
\definevar{demo.education.Master'sorprofessionaldegree.perc.Text}{Zero}
\definevar{demo.education.Somecollege(orcurrentlyenrolled),ortwo-yearassociate’sdegree.perc.text}{zero}
\definevar{demo.education.Somecollege(orcurrentlyenrolled),ortwo-yearassociate’sdegree.perc.Text}{Zero}
\definevar{demo.education.Lessthanhighschool/GCSEorequivalent.perc.text}{zero}
\definevar{demo.education.Lessthanhighschool/GCSEorequivalent.perc.Text}{Zero}
\definevar{demo.education.Somegraduateschool,orcurrentlyenrolledingraduateschool.perc.text}{zero}
\definevar{demo.education.Somegraduateschool,orcurrentlyenrolledingraduateschool.perc.Text}{Zero}
\definevar{demo.webDomainExp.Prefernottodisclose.text}{zero}
\definevar{demo.webDomainExp.Prefernottodisclose.Text}{Zero}
\definevar{demo.webDomainExp.Full-StackDevelopment.text}{15}
\definevar{demo.webDomainExp.Full-StackDevelopment.Text}{15}
\definevar{demo.webDomainExp.FrontendDevelopment.text}{eleven}
\definevar{demo.webDomainExp.FrontendDevelopment.Text}{Eleven}
\definevar{demo.webDomainExp.BackendDevelopment.text}{nine}
\definevar{demo.webDomainExp.BackendDevelopment.Text}{Nine}
\definevar{demo.webDomainExp.WebDesign.text}{seven}
\definevar{demo.webDomainExp.WebDesign.Text}{Seven}
\definevar{demo.webDomainExp.UI/UX.text}{six}
\definevar{demo.webDomainExp.UI/UX.Text}{Six}
\definevar{demo.webDomainExp.APIDevelopment.text}{six}
\definevar{demo.webDomainExp.APIDevelopment.Text}{Six}
\definevar{demo.webDomainExp.Testing.text}{four}
\definevar{demo.webDomainExp.Testing.Text}{Four}
\definevar{demo.webDomainExp.Other(pleasespecifybelow).text}{one}
\definevar{demo.webDomainExp.Other(pleasespecifybelow).Text}{One}
\definevar{demo.webDomainExp.Full-StackDevelopment.perc.text}{zero}
\definevar{demo.webDomainExp.Full-StackDevelopment.perc.Text}{Zero}
\definevar{demo.webDomainExp.FrontendDevelopment.perc.text}{zero}
\definevar{demo.webDomainExp.FrontendDevelopment.perc.Text}{Zero}
\definevar{demo.webDomainExp.BackendDevelopment.perc.text}{zero}
\definevar{demo.webDomainExp.BackendDevelopment.perc.Text}{Zero}
\definevar{demo.webDomainExp.WebDesign.perc.text}{zero}
\definevar{demo.webDomainExp.WebDesign.perc.Text}{Zero}
\definevar{demo.webDomainExp.UI/UX.perc.text}{zero}
\definevar{demo.webDomainExp.UI/UX.perc.Text}{Zero}
\definevar{demo.webDomainExp.APIDevelopment.perc.text}{zero}
\definevar{demo.webDomainExp.APIDevelopment.perc.Text}{Zero}
\definevar{demo.webDomainExp.Testing.perc.text}{zero}
\definevar{demo.webDomainExp.Testing.perc.Text}{Zero}
\definevar{demo.webDomainExp.Other(pleasespecifybelow).perc.text}{zero}
\definevar{demo.webDomainExp.Other(pleasespecifybelow).perc.Text}{Zero}
\definevar{demo.usability.experience.Prefernottodisclose.text}{zero}
\definevar{demo.usability.experience.Prefernottodisclose.Text}{Zero}
\definevar{demo.usability.experience.Some.text}{eight}
\definevar{demo.usability.experience.Some.Text}{Eight}
\definevar{demo.usability.experience.Considerable.text}{seven}
\definevar{demo.usability.experience.Considerable.Text}{Seven}
\definevar{demo.usability.experience.Little.text}{two}
\definevar{demo.usability.experience.Little.Text}{Two}
\definevar{demo.usability.experience.No.text}{one}
\definevar{demo.usability.experience.No.Text}{One}
\definevar{demo.usability.experience.Some.perc.text}{zero}
\definevar{demo.usability.experience.Some.perc.Text}{Zero}
\definevar{demo.usability.experience.Considerable.perc.text}{zero}
\definevar{demo.usability.experience.Considerable.perc.Text}{Zero}
\definevar{demo.usability.experience.Little.perc.text}{zero}
\definevar{demo.usability.experience.Little.perc.Text}{Zero}
\definevar{demo.usability.experience.No.perc.text}{zero}
\definevar{demo.usability.experience.No.perc.Text}{Zero}
\definevar{demo.security.experience.Prefernottodisclose.text}{zero}
\definevar{demo.security.experience.Prefernottodisclose.Text}{Zero}
\definevar{demo.security.experience.Some.text}{12}
\definevar{demo.security.experience.Some.Text}{12}
\definevar{demo.security.experience.Considerable.text}{three}
\definevar{demo.security.experience.Considerable.Text}{Three}
\definevar{demo.security.experience.Little.text}{two}
\definevar{demo.security.experience.Little.Text}{Two}
\definevar{demo.security.experience.No.text}{one}
\definevar{demo.security.experience.No.Text}{One}
\definevar{demo.security.experience.Some.perc.text}{zero}
\definevar{demo.security.experience.Some.perc.Text}{Zero}
\definevar{demo.security.experience.Considerable.perc.text}{zero}
\definevar{demo.security.experience.Considerable.perc.Text}{Zero}
\definevar{demo.security.experience.Little.perc.text}{zero}
\definevar{demo.security.experience.Little.perc.Text}{Zero}
\definevar{demo.security.experience.No.perc.text}{zero}
\definevar{demo.security.experience.No.perc.Text}{Zero}
\definevar{demo.usability.training.studies.Prefernottodisclose.text}{zero}
\definevar{demo.usability.training.studies.Prefernottodisclose.Text}{Zero}
\definevar{demo.usability.training.studies.No.text}{13}
\definevar{demo.usability.training.studies.No.Text}{13}
\definevar{demo.usability.training.studies.Yes.text}{five}
\definevar{demo.usability.training.studies.Yes.Text}{Five}
\definevar{demo.usability.training.studies.No.perc.text}{zero}
\definevar{demo.usability.training.studies.No.perc.Text}{Zero}
\definevar{demo.usability.training.studies.Yes.perc.text}{zero}
\definevar{demo.usability.training.studies.Yes.perc.Text}{Zero}
\definevar{demo.security.training.studies.Prefernottodisclose.text}{zero}
\definevar{demo.security.training.studies.Prefernottodisclose.Text}{Zero}
\definevar{demo.security.training.studies.No.text}{ten}
\definevar{demo.security.training.studies.No.Text}{Ten}
\definevar{demo.security.training.studies.Yes.text}{eight}
\definevar{demo.security.training.studies.Yes.Text}{Eight}
\definevar{demo.security.training.studies.No.perc.text}{zero}
\definevar{demo.security.training.studies.No.perc.Text}{Zero}
\definevar{demo.security.training.studies.Yes.perc.text}{zero}
\definevar{demo.security.training.studies.Yes.perc.Text}{Zero}
\definevar{demo.usability.training.job.Prefernottodisclose.text}{zero}
\definevar{demo.usability.training.job.Prefernottodisclose.Text}{Zero}
\definevar{demo.usability.training.job.No.text}{eleven}
\definevar{demo.usability.training.job.No.Text}{Eleven}
\definevar{demo.usability.training.job.Yes.text}{seven}
\definevar{demo.usability.training.job.Yes.Text}{Seven}
\definevar{demo.usability.training.job.No.perc.text}{zero}
\definevar{demo.usability.training.job.No.perc.Text}{Zero}
\definevar{demo.usability.training.job.Yes.perc.text}{zero}
\definevar{demo.usability.training.job.Yes.perc.Text}{Zero}
\definevar{demo.security.training.job.Prefernottodisclose.text}{zero}
\definevar{demo.security.training.job.Prefernottodisclose.Text}{Zero}
\definevar{demo.security.training.job.No.text}{eleven}
\definevar{demo.security.training.job.No.Text}{Eleven}
\definevar{demo.security.training.job.Yes.text}{seven}
\definevar{demo.security.training.job.Yes.Text}{Seven}
\definevar{demo.security.training.job.No.perc.text}{zero}
\definevar{demo.security.training.job.No.perc.Text}{Zero}
\definevar{demo.security.training.job.Yes.perc.text}{zero}
\definevar{demo.security.training.job.Yes.perc.Text}{Zero}

\definevar{searchengine.Google}{18}
\definevar{searchengine.GoogleScholar}{1}
\definevar{searchengine.Dogpile}{1}
\definevar{searchengine.DuckDuckGo}{3}
\definevar{searchengine.Google.perc}{100.0\%}
\definevar{searchengine.GoogleScholar.perc}{5.6\%}
\definevar{searchengine.Dogpile.perc}{5.6\%}
\definevar{searchengine.DuckDuckGo.perc}{16.7\%}
\definevar{searchengine.Google.text}{18}
\definevar{searchengine.Google.Text}{18}
\definevar{searchengine.GoogleScholar.text}{one}
\definevar{searchengine.GoogleScholar.Text}{One}
\definevar{searchengine.Dogpile.text}{one}
\definevar{searchengine.Dogpile.Text}{One}
\definevar{searchengine.DuckDuckGo.text}{three}
\definevar{searchengine.DuckDuckGo.Text}{Three}
\definevar{searchengine.Google.perc.text}{one}
\definevar{searchengine.Google.perc.Text}{One}
\definevar{searchengine.GoogleScholar.perc.text}{zero}
\definevar{searchengine.GoogleScholar.perc.Text}{Zero}
\definevar{searchengine.Dogpile.perc.text}{zero}
\definevar{searchengine.Dogpile.perc.Text}{Zero}
\definevar{searchengine.DuckDuckGo.perc.text}{zero}
\definevar{searchengine.DuckDuckGo.perc.Text}{Zero}

\definevar{task.documents.upwork}{264}
\definevar{task.documents.pilot}{48}
\definevar{task.queries.upwork}{145}
\definevar{task.queries.pilot}{38}
\definevar{task.queries.all}{183}
\definevar{task.documents.upwork.text}{264}
\definevar{task.documents.upwork.Text}{264}
\definevar{task.documents.pilot.text}{48}
\definevar{task.documents.pilot.Text}{48}
\definevar{task.queries.upwork.text}{145}
\definevar{task.queries.upwork.Text}{145}
\definevar{task.queries.pilot.text}{38}
\definevar{task.queries.pilot.Text}{38}
\definevar{task.queries.all.text}{183}
\definevar{task.queries.all.Text}{183}

\definevar{JointlyCodedDocuments}{10}
\definevar{CodingMergeInterval}{50}
\definevar{SearchExpansionResultsTopN}{10}
\definevar{ParticipantCompensation}{\$45}

\definevar{participants.pilot}{9}
\definevar{participants.pilot.text}{nine}
\definevar{participants.pilot.Text}{Nine}

\definevar{task.documents.searchexpansion}{100}

\definevar{merges}{8}
\definevar{merges.text}{eight}
\definevar{merges.Text}{Eight}
\definevar{merge.criterion}{50}
\definevar{merge.criterion.text}{50}
\definevar{merge.criterion.Text}{50}

\definevar{PersonHours}{232}

\definevar{task.documents.upwork.withduplicates}{284}
\definevar{task.documents.upwork.domains.unique}{171}

\usepackage[acronym, %
    nohypertypes={acronym}, %
]{glossaries}
\newacronym{2fa}{2FA}{two-factor authentication}
\newacronym{mfa}{MFA}{multi-factor authentication}
\newacronym{sso}{SSO}{single sign-on}
\newacronym{ui}{UI}{user interface}
\newacronym{ux}{UX}{user experience}
\newacronym{otp}{OTP}{one-time password}
\newacronym{jwt}{JWT}{JSON Web Token}
\newacronym{irb}{IRB}{institutional review board}
\newacronym{erb}{ERB}{ethical review board}
\newacronym{gdpr}{GDPR}{General Data Protection Regulation}
\newacronym{api}{API}{application programming interface}
\newacronym{ctr}{CTR}{click-through rate}
\newacronym{ai}{AI}{artificial intelligence}
\newacronym{ml}{ML}{machine learning}
\newacronym{hipaa}{HIPAA}{Health Insurance Portability and Accountability Act}
\newacronym{pipeda}{PIPEDA}{Personal Information Protection and Electronic Documents Act}
\newacronym{so}{SO}{Stack Overflow}
\newacronym{spof}{SPOF}{single point of failure}
\newacronym{pwm}{PWM}{password manager}
\newacronym{qda}{QDA}{qualitative data analysis}
\newacronym{irr}{IRR}{inter-rater reliability}

\newacronym{nist}{NIST}{U.S.\ National Institute of Standards and Technology}
\newacronym{enisa}{ENISA}{European Union Agency for Cybersecurity}
\newacronym{bsi}{BSI}{German Federal Office for Information Security}
\newacronym{ncsc}{NCSC}{National Cyber Security Centre} %

\newacronym{owasp}{OWASP}{Open Web Application Security Project}

\extendedversion{}{%
\renewbibmacro{finentry}{%
  \iffieldequalstr{entrykey}{reese2019usability}%
   {\finentry\balance}
   {\finentry}}
}

\copyrightyear{2023}
\acmYear{2023}
\setcopyright{rightsretained}
\acmConference[CCS '23]{Proceedings of the 2023 ACM SIGSAC Conference on Computer and Communications Security}{November 26--30, 2023}{Copenhagen, Denmark}
\acmBooktitle{Proceedings of the 2023 ACM SIGSAC Conference on Computer and Communications Security (CCS '23), November 26--30, 2023, Copenhagen, Denmark}
\acmDOI{10.1145/3576915.3623072}
\acmISBN{979-8-4007-0050-7/23/11}

\makeatletter
\gdef\@copyrightpermission{
  \begin{minipage}{0.3\columnwidth}
   \href{https://creativecommons.org/licenses/by/4.0/}{\includegraphics[width=0.90\textwidth]{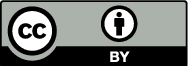}}
  \end{minipage}\hfill
  \begin{minipage}{0.7\columnwidth}
   \href{https://creativecommons.org/licenses/by/4.0/}{This work is licensed under a Creative Commons Attribution International 4.0 License.}
  \end{minipage}
  \vspace{5pt}
}
\makeatother

\usepackage[available]{usenixbadges}

\extendedversion{\usepackage{pdfpages}}{}

\begin{document}

\title[A Qualitative Exploration of Online Developer Advice on Usable and Secure Authentication]{“Make Them Change it Every Week!”: A Qualitative Exploration of Online Developer Advice on Usable and Secure Authentication}
\extendedversion{%
    \subtitle{Extended Version}
    \subtitlenote{%
        This is the extended version of the paper originally published at ACM CCS 2023 (DOI: \href{https://doi.org/10.1145/3576915.3623072}{10.1145/3576915.3623072}). It contains additional information and appendices.
    }
}{
    \titlenote{%
        This paper has an extended version: \url{https://arxiv.org/abs/2309.00744}.
    }
}

\author{Jan H.\ Klemmer}
\orcid{0000-0002-6994-7206}
\affiliation{%
  \institution{Leibniz University Hannover}
  \city{Hannover}
  \country{Germany}
}
\email{klemmer@sec.uni-hannover.de}

\author{Marco Gutfleisch}
\orcid{0000-0003-1400-5825}
\affiliation{%
  \institution{Ruhr University Bochum}
  \city{Bochum}
  \country{Germany}
}
\email{marco.gutfleisch@rub.de}

\author{Christian Stransky}
\orcid{0000-0002-7942-5372}
\affiliation{%
  \institution{CISPA Helmholtz Center for Information Security}
  \city{Hannover}
  \country{Germany}
}
\email{stransky@sec.uni-hannover.de}

\author{Yasemin Acar}
\orcid{0000-0001-7167-7383}
\affiliation{%
 \institution{Paderborn University}
 \city{Paderborn}
 \country{Germany}
}
\email{yasemin.acar@uni-paderborn.de}

\author{M.\ Angela Sasse}
\orcid{0000-0003-1823-5505}
\affiliation{%
  \institution{Ruhr University Bochum}
  \city{Bochum}
  \country{Germany}
}
\email{martina.sasse@rub.de}

\author{Sascha Fahl}
\orcid{0000-0002-5644-3316}
\affiliation{%
  \institution{CISPA Helmholtz Center for Information Security}
  \city{Hannover}
  \country{Germany}
}
\email{sascha.fahl@cispa.de}

\renewcommand{\shortauthors}{Jan H.\ Klemmer et al.}

\ccsdesc[500]{Security and privacy~Usability in security and privacy}
\ccsdesc[300]{Security and privacy~Software security engineering}
\ccsdesc[100]{Security and privacy~Authentication}

\keywords{Usable Security; Advice; Software Development; Authentication}

\begin{abstract}

Usable and secure authentication on the web and beyond is mission-critical. 
While password-based authentication is still widespread, users have trouble dealing with potentially hundreds of online accounts and their passwords. 
Alternatives or extensions such as multi-factor authentication have their own challenges and find only limited adoption. 
Finding the right balance between security and usability is challenging for developers. 
Previous work found that developers use online resources to inform security decisions when writing code. 
Similar to other areas, lots of authentication advice for developers is available online, including blog posts, discussions on Stack Overflow, research papers, or guidelines by institutions like OWASP or NIST.

We are the first to explore developer advice on authentication that affects usable security for end-users.
Based on a survey with \var{participants.Upwork}~professional web developers, we obtained \var{CorpusDocumentCountTotal}~documents and qualitatively analyzed \var{DistinctAdviceCountRelevant}~contained pieces of advice in depth.
We aim to understand the accessibility and quality of online advice and provide insights into how online advice might contribute to (in)secure and (un)usable authentication.
We find that advice is scattered and that finding recommendable, consistent advice is a challenge for developers, among others. 
The most common advice is for password-based authentication, but little for more modern alternatives. 
Unfortunately, many pieces of advice are debatable (\eg{} complex password policies), outdated (\eg{} enforcing regular password changes), or contradicting and might lead to unusable or insecure authentication.
Based on our findings, we make recommendations for developers, advice providers, official institutions, and academia on how to improve online advice for developers.

\end{abstract}

\maketitle

\section{Introduction}
\label{sec:introduction}

Password-based authentication is the status quo on the web and beyond~\cite{Adams:1999uj, Bonneau:2012:sok, Bonneau2015a}---despite its many known problems. This includes challenges around memorizing passwords, password reuse~\cite{Pearman:ccs17:ObservingPasswords, Stobert:2014:PasswordLifecycle, Golla:ccs18:PasswordReuseNotifications}, easily guessable and insecurely stored passwords~\cite{Stobert:2014:PasswordLifecycle, Komanduri:2011gl}, being forced to follow complicated password policies and change passwords regularly~\cite{Lee2022, Komanduri:2011gl, Shay:2014:LongPasswords, Shay:2016:PasswordPolicies}, or phishing attacks~\cite{Lain_2022}. 
Eventually, passwords are a prime example of the importance of \emph{usable} security: 
The underlying problem is not the technical security of passwords but the overwhelming challenges they present to users that negatively impact authentication security.
Replacing passwords with potentially more usable and secure alternatives has been a popular idea for many years~\cite{Bonneau:2012:sok}. 
However, alternatives such as \gls{2fa} and FIDO2 both lack adoption~\cite{farke_you_2020, ghorbani_lyastani_is_2020, conf/ndss/LyastaniBB23, AlQahtani2022, Colnago2018, Petsas2015}, and passwords still remain the \emph{de facto} standard.

Overall, companies, development teams, individual developers, and other involved stakeholders significantly impact the usable security of software systems in general and authentication in particular~\cite{Gutfleisch:2022:usec-in-sdps}. 
For example, they make decisions on password composition policies or the deployment of \gls{2fa}. 
To inform security-related software development decisions, including decisions on authentication deployments, developers draw heavily on online security advice, \eg{} from \gls{so}~\cite{Acar:2016ww, acar:2017:secdev, Acar:2017:InternetResources, Fischer:2017, Fischer:2019:StackOverflowHelpful, Fischer:2021:ContentReranking, Chen:2019:ICSE:HowReliable, Yang:2016:WhatSecurityQuestions, Naiakshina:2017}. 
In the ACM CCS 2022 keynote, Michelle Mazurek concluded that security advice is a \enquote{disaster} and affects developers and other stakeholders~\cite{Mazurek2022}.
As developers draw on security online advice and online advice on authentication is common, the question of what developers can learn from such advice to make authentication deployments usable and secure was the main driver of this work.
Based on a qualitative exploration of the \emph{status quo} of online security advice on authentication deployments for developers, we aimed to develop further ideas for improving future advice.
Previous work on general security advice for end-users~\cite{Hasegawa2022, redmiles_how_2016, fagan_why_2016, Ion2015, busse_replication_2019, Reeder2017, redmiles_comprehensive_2020} and for developers~\cite{Acar:2016ww, acar:2017:secdev, Acar:2017:InternetResources, Fischer:2017, Fischer:2019:StackOverflowHelpful, Fischer:2021:ContentReranking, Chen:2019:ICSE:HowReliable, Yang:2016:WhatSecurityQuestions, Barrera2022, Barrera2022a, Nguyen:2017, Geierhaas2022, Gorski:2018, reeder2011helping} does not address online security advice for developers on implementing usable security for end-users. 
To the best of our knowledge, we are the first to explore this research gap. 

\boldparagraph{Goals}
In this paper, we aim to qualitatively explore developer security advice on implementing authentication that affects end-user usable security.
We aim to understand the accessibility and quality of online advice and to provide insights into how online advice might contribute to (un)usable and (in)secure  authentication.
More specifically, we seek to explore which topics are covered and identify potential challenges developers might face when following advice on implementing authentication online.
Based on our findings, we make recommendations for future advice to better support developers in implementing usable and secure authentication.

\boldparagraph{Approach}
In our qualitative exploratory in-depth analysis, we applied the following multi-step process:
First, we collected online documents that contained advice on authentication. 
Therefore, we recruited \var{participants.Upwork}~professional web developers to collect a set of \var{task.documents.upwork}~online resources (\autoref{sec:methodology-finding-advice}) and their Google search queries. 
We used these queries to enrich the document corpus further. 
The overall document corpus consists of \var{CorpusDocumentCountTotal}~documents on web authentication. 
Second, three researchers extracted all advice on web authentication from each document through qualitative open coding  (\autoref{ssec:extraction-filtering}). 
Finally, in an in-depth qualitative analysis, we explored advice distribution (\autoref{ssec:advice-distribution}) and topics (\autoref{ssec:categorization}) and investigated characteristics like recommendable (\autoref{ssec:recommendable-advice}), debatable (\autoref{ssec:debatable-advice}), and contradicting advice (\autoref{ssec:contradicting-advice}).

\boldparagraph{Key Findings}
The advice we found roughly reflected current (usable) security authentication challenges on the web, \eg{} passwords are still ubiquitous, and password advice was most prevalent (\var{advice.category.passwords.documents.perc} of documents). 
For example, the advice covered password composition policies or password recovery. 
Moreover, only about \var{DistinctAdviceCountRelevantPercent}\% of advice was relevant to usable security of authentication.
We classified \var{advice.recommendable.perc} to be recommendable. 
However, we also found \var{advice.questionable.perc} to be debatable, \eg{} being outdated, factually wrong, or counteracting usable security (\eg{} enforcing regular password changes). 
We identified \var{advice.contradicting.groups}~groups of two or more pieces of advice that contradict each other. 
Overall, we found that advice from research results and official institutions was rarely integrated into the advice we analyzed.
Furthermore, the advice was scattered across many online resources, making it difficult for developers to find. 

\boldparagraph{Contributions}
In this paper, we make the following contributions:
(i)~To the best of our knowledge, this is the first study investigating online advice for developers on authentication that affects security and usability for end-users. 
(ii)~We present an overview of authentication advice topics and further explore recommendable, debatable, and contradicting advice. 
(iii)~Based on our findings, we make recommendations for future usable security advice for developers. 
Furthermore, we highlight open challenges and outline future research. 
(iv)~Finally, we provide our document corpus, all pieces of advice, and additional supplementary material in a replication package (\autoref{sec:replication-package}).

\section{Related Work}
\label{sec:rel-work}

We discuss related work in three key areas: 
(i)~general security advice for end-users, 
(ii)~(usable) security advice for developers, 
and (iii)~usable and secure web authentication and its challenges.
We highlight our novel contributions and put it into context.

\boldparagraph{General End-User Security Advice}
A large portion of online advice on security addresses end-users~\cite{redmiles_comprehensive_2020, redmiles_how_2016, fagan_why_2016, Ion2015, busse_replication_2019, conf/ssp/MunyendoAA23, conf/nspw/Herley09}, \eg{} concerning topics like how to handle passwords or spam. 
Non-experts, like most end-users, heavily request such advice~\cite{Hasegawa2022}. 
As lots of advice exists, especially online, the selection of advice sources has to be considered~\cite{redmiles_how_2016}. 
For example, factors like the source's perceived trustworthiness~\cite{redmiles_how_2016} and convenience~\cite{fagan_why_2016} influence advice adoption.
One explanation for low advice adoption is that users rationally reject advice as it often offers a poor cost--benefit tradeoff~\cite{conf/nspw/Herley09}.
Overall, security advice can be challenging to adopt, as a mismatch between advice and behavior was perceived among both experts and non-experts~\cite{Ion2015, busse_replication_2019}.
Even security experts were found to lack consensus for advice~\cite{Reeder2017}. 
Similarly, users and experts (like administrators) struggle with advice prioritization~\cite{redmiles_comprehensive_2020}.

While advice can be used to educate end-users, we argue that it is (at least equally) important to help software professionals design systems that people can use securely---ideally without users needing any advice. 
Therefore, we focus on advice for software developers. %

\boldparagraph{(Usable) Security Advice for Developers}
In recent years, researching human factors and security with a focus on developers has gained popularity~\cite{Acar:2016:SecDev, Green:DevNot, Pieczul2017DevelopercenteredSA, tahaei_survey_2019}. This is also true for research on security advice for developers. 
Several studies found that advice impacts security and that sources containing insecure code find their way into software~\cite{Acar:2016ww, acar:2017:secdev, Acar:2017:InternetResources}. For example, Acar et al.\ found that developers who used \gls{so} produced significantly less secure code and that only 17\% of \gls{so} posts contained secure code snippets. Insecure code from \gls{so} is also prevalent in many Android apps~\cite{Fischer:2017} and contained in top Google search results~\cite{Fischer:2021:ContentReranking}. 
However, nudging~\cite{Fischer:2019:StackOverflowHelpful} and re-ranking search results~\cite{Fischer:2021:ContentReranking} can create a positive security impact of \gls{so}. 
Overall, \gls{so} is a place for software professionals to discuss security-relevant topics~\cite{Yang:2016:WhatSecurityQuestions}, while secure and insecure answers were almost balanced~\cite{Chen:2019:ICSE:HowReliable}.
Apart from \gls{so}, researchers also found challenges in other advice sources, \eg{} not actionable IoT advice~\cite{Barrera2022, Barrera2022a}.

Advice can also be given through other channels. This includes IDE plug-ins that aid developers in preventing security pitfalls. For example, Nguyen et al.\ presented and evaluated a plug-in that supports Android developers~\cite{Nguyen:2017}. Similar tools exist for other use cases, \eg{} secure password storage~\cite{Geierhaas2022}. Besides such tools, Gorski et al.\ found that integrating security advice into \glspl{api} can significantly improve code security without affecting \gls{api} usability~\cite{Gorski:2018}.
Other advice forms can be more general by outlining principles (\eg{} Microsoft's \emph{NEAT} design principles for security warnings~\cite{reeder2011helping}) and abstract guidance (\eg{} heuristics for accessibility and usable security on the web~\cite{napoli_developing_2018}).

Most similar to our work, Iacono, Gorski, et al.\ systematized security principles, guidelines, and patterns based on scientific literature~\cite{iacono_consolidating_2018, gorski_providing_2019}, but only \enquote{concentrated on principles and patterns for end-users}~\cite{gorski_providing_2019}. 
In contrast, we focus especially on advice for developers that affects end-users' usable security. Moreover, we focus specifically on (non-scientific) online resources.
To the best of our knowledge, no paper examined advice for developers that covers end-user usable security. We argue that end-user (usable) security, however, highly depends on decisions and implementation during software development. Our analysis addresses this gap.

\boldparagraph{Usable and Secure Web Authentication}

Passwords, probably the oldest authentication approach~\cite{Bonneau2015a} and a starting point for early usable security research~\cite{Adams:1999uj}, are the most used authentication approach despite many shortcomings.
This includes but is not limited to insecure storage~\cite{Stobert:2014:PasswordLifecycle, Komanduri:2011gl}, password reuse~\cite{Pearman:ccs17:ObservingPasswords, Stobert:2014:PasswordLifecycle, Golla:ccs18:PasswordReuseNotifications}, and usability and (resulting) security problems through password policies~\cite{Lee2022, Komanduri:2011gl, Shay:2014:LongPasswords, Shay:2016:PasswordPolicies, Shay:2012:SystemAssignedPassphrases}.
To aid in creating secure passwords, password strength meters are a common tool and visualization technique~\cite{Ur:2012ug}. However, their construction is difficult. For example, researchers found inconsistencies, weaknesses~\cite{Carnavalet:2014:PasswordMeters}, and accuracy issues~\cite{golla_accuracy_2018, conf/sec/WangSDSJ23}.
Similarly, \glspl{pwm} should help users deal with their passwords, \eg{} by creating secure (random) passwords~\cite{Zibaei2022} that fit websites' password policies~\cite{Gautam2022}. However, even \glspl{pwm} have interaction problems with websites~\cite{Huaman:2021:pwms} or overwhelm their users~\cite{conf/chi/OeaschRSG22}, both limiting their usability. 
Due to passwords' numerous problems, the idea of replacing them became popular. 
However, many approaches failed~\cite{Bonneau:2012:sok}, and replacing passwords remains an unsolved challenge.

For example, replacing passwords could be achieved with the FIDO2 standard~\cite{FIDO2}. 
However, FIDO2 lacks adoption, which might be caused by overlooked benefits~\cite{farke_you_2020}, concerns, and not trusting the new technology despite good usability~\cite{ghorbani_lyastani_is_2020}. 
Despite generally good usability, users struggle with hardware token setup~\cite{reynolds_tale_2018, Das2018} or rate the usability of smartphones for FIDO2 as low~\cite{Owens2021}. 
As tokens can get lost and are associated with costs, electronic IDs can be used for FIDO2~\cite{Schwarz2022}---albeit being hard to set up~\cite{Keil2022}.

\Gls{2fa}, or generally \gls{mfa}, is an important concept for strengthening security---especially when using passwords---but could also negatively affect usability~\cite{reynolds_empirical_2020, reese2019usability, Abbott2020}. 
Users might have problems with the initial setup~\cite{reese2019usability}, and the \gls{2fa} \gls{ux} is inconsistent~\cite{conf/ndss/LyastaniBB23}. 
Overall, \gls{2fa} adoption is still relatively low and needs to be increased~\cite{Redmiles:2017:2FAMessages, AlQahtani2022, Colnago2018, Petsas2015}.
Widely adopted approaches like SMS-based \gls{2fa} are known to be vulnerable to phishing~\cite{markert_view_2019}.

To conclude, authentication has pitfalls that software developers need to consider. 
Interestingly, most outlined challenges are no technical security problems and boil down to a usability problem. For example, passwords would be more secure if used correctly. %
Therefore, we investigate how developer advice supports usable and secure authentication.

\section{Collecting Online Developer Advice}
\label{sec:methodology-finding-advice}
In this section, we describe our data collection of \var{CorpusDocumentCountTotal}~online resources on authentication. 
While we aimed to create an ecologically valid dataset, our main goal was diversity due to the exploratory nature of our work.
In a developer survey between August and November 2021, \var{participants.Upwork.text}~professional web developers searched and reported online resources on web authentication they considered for an imaginary web authentication development task. 
We extend the document corpus further by including search results for our participants' search queries. 
As we had already reached saturation with this document corpus, we stopped recruiting more participants.

\subsection{Developer Survey}
\label{ssec:dev-survey}
Our goal was to create a comprehensive corpus of documents containing online advice regarding usable and secure web authentication for developers that impacts usable security for end-users. 
We asked professional software developers to search for online resources they consider relevant for implementing usable and secure web authentication. 
We gave them a scenario and task to search for relevant advice on the web.

\subsubsection{Scenario}
We created a task containing a web development scenario and framed it towards authentication and usable security. 
We aimed to collect document URLs, search queries, and other information from software developers.\extendedversion{\footnote{See Appendix~\ref{sec:appendix-scenario-task} for the full scenario description and search task.}}{\footnote{See this paper's extended version for the full scenario description and search task.}}
Our scenario consisted of a hypothetical new web-based health platform \emph{KeepYourHealth}, that should become the central pivot point for personal health, \eg{} storing and sharing medical data or communicating with doctors.
We chose the healthcare framing for three reasons: 
First, medical and personal health information is considered highly sensitive, hopefully shifting the participants' attention to security. 
Second, to highlight the need for good usability, medical applications have a diverse user base that likely contains many non-tech-savvy users.
Third, medical applications must follow certain compliance or data protection rules (\eg{} \acrshort{hipaa} in the US, \acrshort{pipeda} in Canada,  \acrshort{gdpr} in Europe). 
While this might skew the results slightly towards official documents, we also hoped to receive a more diverse and complete document sample.

As part of the KeepYourHealth project, we wanted participants to imagine that they are responsible for designing and implementing the authentication mechanism for both a mobile and a desktop version. 
To create the concept for KeepYourHealth's authentication system (including registration and login), the participants had to perform online research and look for helpful resources. 
We designed the scenario to be very open to ensure diverse results, as participants can explore various topics in their search. 
Moreover, we instructed participants to freely consider any resource, authentication approach, technology, etc. 
Our only requirement was that the authentication system had to be usable and secure.
We used words like \emph{easy-to-use}, \emph{user-friendly}, \emph{secure}, \emph{usability}, and \emph{security} to emphasize the requirement for usable security.

\subsubsection{Search Task}
To collect online advice, we gave software professionals the task of searching for online resources that provide advice that potentially helps to build a usable and secure web authentication system for KeepYourHealth. 
We explained our research project's purpose and that we wanted to collect and analyze advice from various online sources. 
We additionally explained that we, as researchers, are biased and probably would search differently than web development experts. 
We also wanted the software developers to report how they found a specific online source, \ie{} the search query. 
We provided the participants with a table to report documents and search queries. The table consisted of the following columns\extendedversion{ (cf.\ \autoref{fig:screenshot-search-task-table})}{}: URL, how the document was found (\eg{} bookmark, link within another site, search engine), and (if applicable) the search query that yielded the document. 
Each column's purpose was also explained.
Our participants were free to choose any search approach they wanted, \ie{} we did not constrain them to a specific search engine. 

\subsubsection{Questionnaire}
We embedded the search task in an online questionnaire. 
After an initial consent form, the participants were introduced to the scenario and their tasks. 
We gave them 30~minutes for the task. 
For the search task, we displayed the above-mentioned table for the users to enter URLs and search queries.
After completing the task, we asked the participants for their most used search engine and to rank different information resource types by their helpfulness. 
Next, we followed up with standard demographic questions, including age, gender, country of residence, highest level of education, and employment status. 
Additionally, we asked some questions specific to this study's context regarding participants' experience, skills, and security and usability education. 
We estimated the maximum time for the whole survey at 45~minutes. 
The replication package (Appendix~\ref{sec:replication-package}) includes the full questionnaire.

\subsubsection{Instrument Development \& Piloting}
We developed and piloted the questionnaire, scenario, and task in multiple iterations. 
We asked \var{participants.pilot.text}~persons to pilot the study and questionnaire, including researchers with experience conducting usable security studies and experiments. 
We made smaller adjustments regarding wording and description texts based on the feedback that we collected via think-aloud and feedback text boxes on each survey page. 
Finally, we tested the whole setup with three professional developers until no further adjustments were necessary. 
During piloting, we collected \var{task.documents.pilot}~documents and \var{task.queries.pilot}~search queries. 
We included those in our final document corpus to enhance its diversity further.

\subsubsection{Ethics \& Data Protection}
We received \gls{erb} approval for this study from one of the involved institutions, concluding that there were no ethical concerns. 
The remaining institutions had neither an \gls{irb} nor an \gls{erb} that applied to our study. 
This study was designed to adhere to the ethical research principles of the \emph{Menlo Report}~\cite{Kenneally:2012:menlo}.
We further adhered to the strict German and EU privacy laws. 
All participants consented to collecting, storing, and analyzing their data for this research. 
We clarified that we would only evaluate and publish de-identified and aggregated data. 
Furthermore, we informed all participants that they could terminate the study at any time. %

\subsubsection{Recruiting}
We aimed to recruit a diverse sample of software developers with web development experience. 
Therefore, we used the freelancer platform Upwork. 
We created an Upwork job post describing our study and communicated the idea of our research. 
Participants were required to have sufficient proficiency in English.
Besides that, we checked the Upworkers' profiles and job history to ensure they have web development experience.
We also directly invited Upworkers who had indicated experience in web development in their profiles. 
The job was rewarded with \var{ParticipantCompensation} (hourly wage of \$60), above Upwork's \$15--\$30 median hourly pay for software developers~\cite{upwork-costs} to offer competitive compensation.

\subsubsection{Demographics}
We recruited \var{participants.Upwork.text}~web developers via Upwork whose demographics are comparable with those of professional developers from \gls{so}'s latest developer survey~\cite{2022StackOverflowSurvey}. 
An overview of participants' demographics can be found in \autoref{tab:demographics}. 
The participants were from \var{demo.countries.distinct}~different countries, with the UK and the U.S.\ being the most common, including others such as Canada, Nepal, Romania, Germany, Nigeria, France, and Turkey. The majority had at least a bachelor's degree.
Their professional software development experience varied between \var{demo.exp.min} and \var{demo.exp.max}~years. 
Participants covered various roles, such as software engineers, web developers, architects, team leads, and CTOs. 
Most participants reported some experience with usable security.
Some participants also received education in this area in their training, studies, or professional career.
\begin{table}[t]
    \caption{Demographics of the \var{participants.Upwork.text}~web developers.}
    \label{tab:demographics}
    \centering
    \begin{threeparttable}
        \scriptsize
        \renewcommand{\arraystretch}{1.05}
        \setlength{\tabcolsep}{1.333\tabcolsep}
        \setlength{\defaultaddspace}{0.08\defaultaddspace} %
    
        \begin{tabular}{llrr@{\hspace{2em}}lrr}
            \toprule
            
            \multicolumn{7}{l}{\textbf{Gender}}  \\
             & Male & \var{demo.gender.Man} & \var{demo.gender.Man.perc} & Female & \var{demo.gender.Woman} & \var{demo.gender.Woman.perc} \\
            
            \addlinespace
            \multicolumn{7}{l}{\textbf{Age [years]}} \\
             & Min. & \var{demo.age.min} &  & Max. & \var{demo.age.max} & \\
             & Mean (Std.) & \var{demo.age.mean} & $\pm\var{demo.age.std}$ & Median & \var{demo.age.50percentile}  \\
            
            \addlinespace
            \multicolumn{7}{l}{\textbf{Highest Education Level}}  \\
             & < High school & \var{demo.education.Lessthanhighschool/GCSEorequivalent} & \var{demo.education.Lessthanhighschool/GCSEorequivalent.perc} & Bachelor's degree\hspace{-3em} & \var{demo.education.Bachelor'sdegree} & \var{demo.education.Bachelor'sdegree.perc} \\
             & High school                   & \var{demo.education.Highschoolorequivalent/Alevelorequivalent} & \var{demo.education.Highschoolorequivalent/Alevelorequivalent.perc} & Attending grad.\ \rlap{school} & \var{demo.education.Somegraduateschool,orcurrentlyenrolledingraduateschool} & \var{demo.education.Somegraduateschool,orcurrentlyenrolledingraduateschool.perc} \\
             & College & \var{demo.education.Somecollege(orcurrentlyenrolled),ortwo-yearassociate’sdegree} & \var{demo.education.Somecollege(orcurrentlyenrolled),ortwo-yearassociate’sdegree.perc} & Master's degree & \var{demo.education.Master'sorprofessionaldegree} & \var{demo.education.Master'sorprofessionaldegree.perc} \\
            
            \addlinespace
            \multicolumn{7}{l}{\textbf{Country of Residency}}  \\
            & United Kingdom & \var{demo.countries.UnitedKingdom} & \var{demo.countries.UnitedKingdom.perc} & Brazil & \var{demo.countries.Brazil} & \var{demo.countries.Brazil.perc} \\
            & United States & \var{demo.countries.UnitedStates} & \var{demo.countries.UnitedStates.perc} & Other\tnote{1} & \var{demo.countries.other} & \var{demo.countries.other.perc} \\
            & India & \var{demo.countries.India} & \var{demo.countries.India.perc} \\
            
            \addlinespace
            \multicolumn{7}{l}{\textbf{Industry Experience [years]}}\\ 
            & Min. & \var{demo.exp.min} &  & Max. & \var{demo.exp.max} & \\
            & Mean (Std.) & \var{demo.exp.mean} & $\pm\var{demo.exp.std}$ & Median & \var{demo.exp.50percentile}  \\
            
            \addlinespace
            \multicolumn{7}{l}{\textbf{Employment Status\tnote{\textdagger}}}\\ 
             & Employed full-time & \var{demo.employment.Employedfull-time} & \var{demo.employment.Employedfull-time.perc} & Student & \var{demo.employment.Student} & \var{demo.employment.Student.perc} \\
             & Self-employed/Freelancer\hspace{-1em} & \var{demo.employment.Self-employed/Freelancer} & \var{demo.employment.Self-employed/Freelancer.perc} & Self-describe & \var{demo.employment.Prefertoself-describe} & \var{demo.employment.Prefertoself-describe.perc} \\
             & Employed part-time & \var{demo.employment.Employedpart-time} & \var{demo.employment.Employedpart-time.perc} &  &  &  \\
            
            \addlinespace
            \multicolumn{7}{l}{\textbf{Domain of Web Experience\tnote{\textdagger}}}\\ 
             & Full-stack developer & \var{demo.webDomainExp.Full-StackDevelopment} & \var{demo.webDomainExp.Full-StackDevelopment.perc} & UI/UX & \var{demo.webDomainExp.UI/UX} & \var{demo.webDomainExp.UI/UX.perc} \\
             & Frontend developer & \var{demo.webDomainExp.FrontendDevelopment} & \var{demo.webDomainExp.FrontendDevelopment.perc} & API developer & \var{demo.webDomainExp.APIDevelopment} & \var{demo.webDomainExp.APIDevelopment.perc} \\
             & Backend developer & \var{demo.webDomainExp.BackendDevelopment} & \var{demo.webDomainExp.BackendDevelopment.perc} & Testing & \var{demo.webDomainExp.Testing} & \var{demo.webDomainExp.Testing.perc} \\
             & Web design & \var{demo.webDomainExp.WebDesign} & \var{demo.webDomainExp.WebDesign.perc} & Other & \var{demo.webDomainExp.Other(pleasespecifybelow)} & \var{demo.webDomainExp.Other(pleasespecifybelow).perc} \\
            
            \addlinespace
            \multicolumn{7}{l}{\textbf{Security/Usability as Part of Training/Studies}}\\ 
             & Security & \var{demo.security.training.studies.Yes} & \var{demo.security.training.studies.Yes.perc} & Usability & \var{demo.usability.training.studies.Yes} & \var{demo.usability.training.studies.Yes.perc} \\
            
            \addlinespace
            \multicolumn{7}{l}{\textbf{Security/Usability Training in Professional Activity}}\\ 
             & Security & \var{demo.security.training.job.Yes} & \var{demo.security.training.job.Yes.perc} & Usability & \var{demo.usability.training.job.Yes} & \var{demo.usability.training.job.Yes.perc} \\

            \bottomrule
        \end{tabular}
        \begin{tablenotes}
            \footnotesize
            \item [\textdagger] Multiple answers allowed; may not sum to 100\%.
            \item [1] Each country occurring once.
        \end{tablenotes}
    \end{threeparttable}%
\end{table}

\subsubsection{Reported Documents and Search Queries}
The participants overall found \var{task.documents.upwork}~distinct documents (\var{task.documents.upwork.withduplicates} with duplicates reported by multiple participants) across \var{task.documents.upwork.domains.unique}~different website domains. 
This already indicates that advice is contained in many resources from different authors. 
However, some domains were reported more frequently, as shown in \autoref{tab:most-frequent-domains}. 
This mainly includes authentication platforms like \emph{Auth0}, \emph{Okta}, \emph{swoopnow.com}, but also blogs and tutorial sites like \emph{medium.com} and \emph{dzone.com}, or documentation like the one provided by \emph{Google} or the library \emph{next-auth}. 
The \gls{owasp}'s \emph{Cheat Sheet Series} is included.

We collected \var{task.queries.upwork}~distinct search queries from our participants. 
All participants reported having used Google for their web searches. 
The most frequent terms and term bigrams are shown in Tables~\ref{tab:most-frequent-terms} and~\ref{tab:most-frequent-term-bigrams}. 
While \emph{authentication} is the most frequent term, participants searched specifically for security aspects, \emph{best practices}, and usability aspects.
This also indicates that our task worked as intended, as the terms reflect the goal of usable and secure authentication.

\begin{table*}[tbh]
    \caption{Most frequent domains and search query terms and bigrams that participants reported.}
    \begin{subtable}[c]{0.3\linewidth}
        \centering
        \subcaption{Top-10 documents domains.}
        \label{tab:most-frequent-domains}
        \scriptsize
        \renewcommand{\arraystretch}{1.00}
        \setlength{\tabcolsep}{0.75\tabcolsep}
        \setlength{\defaultaddspace}{0.5\defaultaddspace} %
        \rowcolors{2}{white}{gray!10}
        \begin{tabularx}{\textwidth}{Xr}
            \toprule
                \textbf{Domain} & \textbf{\#} \\
            \midrule
                auth0.com & 23\\
                medium.com & 16\\
                www.okta.com & 8\\
                swoopnow.com & 6\\
                blog.jscrambler.com & 6\\
                cheatsheetseries.owasp.org & 5\\
                cloud.google.com & 5\\
                dzone.com & 5\\
                developers.google.com & 4\\
                next-auth.js.org & 4\\
            \bottomrule
        \end{tabularx}
    \end{subtable}
    \hfill
    \begin{subtable}[c]{0.3\linewidth}
        \centering
        \subcaption{Top-10 terms from search queries.}
        \label{tab:most-frequent-terms}
        \scriptsize
        \renewcommand{\arraystretch}{1.00}
        \setlength{\tabcolsep}{0.75\tabcolsep}
        \setlength{\defaultaddspace}{0.5\defaultaddspace} %
        \rowcolors{2}{white}{gray!10}
        \begin{tabularx}{\textwidth}{Xr}
            \toprule
                \textbf{Search Term} & \textbf{\#} \\
            \midrule
                authentication & 72 \\
                web & 39 \\
                best & 15 \\
                security & 11 \\
                secure & 11 \\
                methods & 10 \\
                user & 10 \\
                app & 9 \\
                practices & 8 \\
                applications & 7 \\
            \bottomrule
        \end{tabularx}
    \end{subtable}
    \hfill
    \begin{subtable}[c]{0.3\linewidth}
        \centering
        \subcaption{Top-10 bigrams from search queries.}
        \label{tab:most-frequent-term-bigrams}
        \scriptsize
        \renewcommand{\arraystretch}{1.00}
        \setlength{\tabcolsep}{0.75\tabcolsep}
        \setlength{\defaultaddspace}{0.5\defaultaddspace} %
        \rowcolors{2}{white}{gray!10}
        \begin{tabularx}{\textwidth}{Xr}
            \toprule
                \textbf{Search Term Bigram} & \textbf{\#} \\
            \midrule
                web authentication & 14 \\
                authentication methods & 10 \\
                for web & 8 \\
                best practices & 8 \\
                web app & 8 \\
                web applications & 7 \\
                user authentication & 6 \\
                authentication best & 5 \\
                methods for & 4 \\
                secure authentication & 4 \\
            \bottomrule
        \end{tabularx}
    \end{subtable}
\end{table*}

\begin{summaryBox}{Summary: Initial Document Collection}
    The search task with \var{participants.Upwork.text}~professional web developers resulted in \var{task.documents.upwork}~documents and \var{task.queries.upwork}~search queries. 
    Both document domains and search queries show high diversity.
\end{summaryBox}

\subsection{Search Expansion}
\label{ssec:document-collection}

We extended our document corpus with search results for the collected search queries that participants reported in addition to the directly reported documents. 
We performed this search expansion because participants might have missed relevant documents to counterbalance participants' personalized search results.
Moreover, this increases overall diversity and helps to reach saturation in the best case, while adding no additional insights in the worst case.
For each query, we obtained the corresponding search results to expand our document corpus. 
Therefore, we leveraged Google's \emph{Custom Search API}~\cite{GoogleProgrammableSearchEngine} to execute a search for each query and extracted its top~\var{SearchExpansionResultsTopN}~results, \ie{} its URL and some metadata. 
We chose this Google-based engine, as Google is the most used search engine~\cite{statcounter:2021:SearchEngines, NetMarketshare:2021:SearchEngines} and all participants reported that they use Google as one of their main search engines. 
We chose \var{SearchExpansionResultsTopN}~results per query, as this corresponds to the first page of search results. 
Moreover, the \gls{ctr} drops below 10\% after the fourth and quickly decreases further to almost 0\% for lower-ranked results~\cite{pan2007google}. 
That said, almost no user looks at the second page of results~\cite{Backlinko:2019:SearchEngineCTR}. 
A~top-\var{SearchExpansionResultsTopN} expansion, therefore, covers search results a user potentially clicks while also including a safety margin.

\subsection{Final Document Collection} 
\label{ssec:finaldocumentcorpus}
The search task with professional software developers resulted in \var{task.documents.upwork}~distinct documents and \var{task.queries.upwork}~distinct search queries. 
As our goal for this qualitative study is diversity, we decided also to include all \var{task.documents.pilot}~documents and \var{task.queries.pilot}~search queries from piloting. %
As described above, we extend our corpus with documents found using Google. 
We randomly sampled \var{task.documents.searchexpansion}~search results from the search expansion and included those in the final document corpus.
Combining all these documents, we obtained a final dataset of \var{CorpusDocumentCountTotal}~documents overall.
For the documents that indicate a publishing date, the median year for the last update was 2021 (oldest: 2002, newest: 2021).
We downloaded each as a PDF for our analysis.

We stopped after these \var{CorpusDocumentCountTotal}~documents, as we had already reached saturation, which means that we would not gain new insights from analyzing more documents and advice, \ie{} no new themes emerged in \autoref{sec:analysis} (\eg{} advice topics). 
Repeated advice in multiple documents also indicated saturation. 
Consequently, we did not survey further participants.

\begin{figure}[t]
    \centering
    \includegraphics[width=0.8\linewidth]{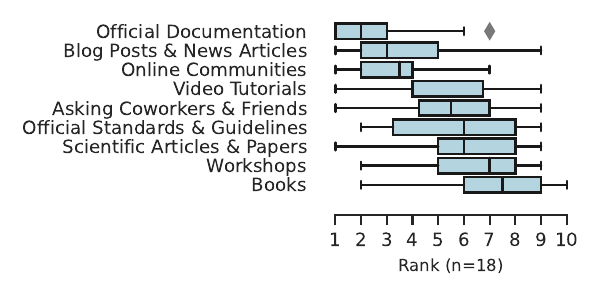}
    \caption{Participants' ranking of information sources for the question: \enquote{Which advice or information sources do you find most helpful in general?}. From \emph{most helpful}~(1) to \emph{least helpful}~(10), excluding ``Other'' answers.}
    \label{fig:info-sources-helpfulness-ranking}
\end{figure}

In the survey, we asked our participants about the helpfulness of different information sources (cf.\ \autoref{fig:info-sources-helpfulness-ranking}). 
Overall, participants reported a high affinity towards online resources. 
The top-three sources were official documentation, blog/news articles, and online communities.
Similarly, this is also reflected in the different types of documents (\autoref{tab:document-types}). 
We found blog posts (some including news) to be the most common type, followed by documentation and wikis and project websites (including source code repositories).
Surprisingly, only \var{documents.type.forum/q&a}~documents (\var{documents.type.forum/q&a.perc}) were from Q\&A forums, 13 of them from \gls{so} (cf.\ discussion in \autoref{ssec:discussion-comparison-rel-work}).
Minor document types included educational material and courses as well as video and audio resources.
\var{documents.type.nothing}~documents had no clear type. 
\var{documents.type.notfound} (\var{documents.type.notfound.perc}) were not available online anymore when we downloaded the documents.

\begin{table}[tb]
    \centering
    \caption{Types of the reported documents.}
    \label{tab:document-types}
    \scriptsize
    \renewcommand{\arraystretch}{1.00}
    \setlength{\tabcolsep}{1\tabcolsep}
    \setlength{\defaultaddspace}{0.5\defaultaddspace} %
    \rowcolors{2}{white}{gray!10}
    \begin{tabular}{lrr}
        \toprule
        \textbf{Type} & \textbf{\#} & \textbf{\%} \\ 
        \midrule
        Blogs (including news) & \var{documents.type.blog/news} & \var{documents.type.blog/news.perc} \\
        Documentation \& Wikis & \var{documents.type.docu/wiki} & \var{documents.type.docu/wiki.perc} \\
        Project Websites \& Code Repositories & \var{documents.type.projectwebsite/repo} & \var{documents.type.projectwebsite/repo.perc} \\
        Scientific & \var{documents.type.paper/scientific} & \var{documents.type.paper/scientific.perc} \\
        Q\&A Forums & \var{documents.type.forum/q&a} & \var{documents.type.forum/q&a.perc} \\
        Educational & \var{documents.type.course/education} & \var{documents.type.course/education.perc} \\
        Video/Audio & \var{documents.type.video/audio} & \var{documents.type.video/audio.perc} \\
        Undefined Type  & \var{documents.type.nothing} & \var{documents.type.nothing.perc} \\
        Not Available & \var{documents.type.notfound} & \var{documents.type.notfound.perc} \\
        \bottomrule
    \end{tabular}
\end{table}

\begin{summaryBox}{Summary: Document Corpus}
    Our final document corpus consisted of \var{CorpusDocumentCountTotal}~distinct documents which we downloaded for analysis. 
    Blog posts, documentation, and wiki pages were the most common document types.
\end{summaryBox}

\subsection{Limitations}
\label{ssec:limitations}

This work has some limitations that are typical for this type of study. 
Generally, the developer survey is subject to self-reporting, social desirability, under- and over-reporting, and sampling biases.
We note that our document corpus cannot contain all online resources and that it is influenced by the participants' experience and especially the personalization of their search results. 
We addressed this by recruiting a diverse set of participants and extending the document corpus with search results until we reached saturation. 
Furthermore, we received only English documents as it was the study's language. 
We decided that English as the only language is an acceptable trade-off as it is the main language used in software development (\eg{} documentation or \gls{so}). 
The search task and scenario (\autoref{ssec:dev-survey}) itself could bias the document corpus. 
However, we favored a realistic scenario without many requirements for external validity over a general, undirected search task.

\section{Advice Analysis}
\label{sec:analysis}

In this section, we provide details on the advice analysis and its results. 
Our analysis is a multi-stage process that starts with identifying and extracting advice (\autoref{ssec:extraction-filtering}) before analyzing it in detail.
In the following subsections, we describe the advice distribution (\autoref{ssec:advice-distribution}), its topics (\autoref{ssec:categorization}), and occurrence of recommendable (\autoref{ssec:recommendable-advice}), debatable (\autoref{ssec:debatable-advice}), and contradicting advice (\autoref{ssec:contradicting-advice}).

Throughout the paper, we report counts, \eg{} counts of advice or documents. As this is qualitative work, those counts are used to convey weight and should \emph{not} be interpreted as quantitative statistical results.
While we try to give an overview by discussing advice examples, we cannot describe and discuss every single piece of advice due to their high number.

\extendedversion{%
\subsection{Positionality Statement}
\label{ssec:positionality}

Before giving detailed insights, we want to highlight that this is qualitative research.
While we always involved multiple researchers and based decisions on external references where possible, this study's design and results might be influenced by the involved researchers.
That said, the results are not guaranteed to be entirely objective and should be interpreted in context.
Therefore, we aim to clarify our backgrounds with this positionality statement. 
All authors are usable security researchers with varying levels of experience, ranging from third-year Ph.D.\ student, to tenured faculty and full professors with more than 30~years usable security research experience.
Our study backgrounds cover computer science, IT security, mathematics, and psychology.
The authors are based in Germany.
}{}

\subsection{Identifying \& Extracting Advice}
\label{ssec:extraction-filtering}

Below, we describe our advice identification and extraction approach in detail. 
It consists of two steps:  
(1)~First, we extracted and systematized each piece of advice from all documents through qualitative coding.
(2)~Second, we applied inclusion and exclusion criteria to reduce the set of advice to address usable and secure web authentication relevant to our study.

\subsubsection{Advice Extraction}

We extracted each advice for developers by reading all documents and manually assigning codes that summarize each piece of advice. 
We used the following definitions to detect advice. 
The definitions' goal was to make the extraction process transparent, objective, and reproducible.
\begin{itemize}
    \item \emph{(Concrete) Advice:} We considered explicit prompts to do something in a specific way or specific recommendations concrete advice. 
    Phrases like ``we recommend,'' ``a developer should,'' or imperatives like ``ensure'' often indicate advice, which the researchers therefore used as proxies to detect advice. 
    An example of such a concrete recommendation is: \enquote{Allow users to toggle password visibility when typing it.}

    \item \emph{Descriptive Information:} We deliberately refrained from extracting information as advice that was not explicitly directed at the reader or any other form of specific recommended action. 
    Some document passages just inform and describe, leave room for interpretation, and require reading between the lines. 
    It might be unclear whether the resource's author wanted to give advice, whether a reader perceives it to be advice, and how one might interpret it.
\end{itemize} 
Three researchers each annotated a third of the final dataset in an iterative open coding process. 
As advice in the documents can be long and wordy, we summarized it by creating codes that briefly paraphrase the advice~\cite{Mayring2015, Kuckartz2019}. 
\autoref{tab:advice-paraphrases} shows examples of original document excerpts and our assigned advice codes. 
As shown in the table, different text passages can have the same meaning and were therefore assigned the same advice code (examples 3 and 4).
We used these advice codes for our analyses throughout the paper.
Each researcher grouped similar or related paraphrases in a code system. 
The leaf nodes in the code tree contain the advice. 
The inner nodes are used to build a hierarchy (\eg{} group all advice on passwords). 
After coding sets of \var{merge.criterion.text}~documents, at least two coders merged, refined, and regrouped the code system if necessary; all three reviewed the resulting code system. 
The coders repeated the described process \var{merges.text}~times until advice from all \var{CorpusDocumentCountTotal}~documents was extracted. 
In accordance with McDonald et al.~\cite{McDonald2019}, we refrain from reporting \gls{irr} because extracting only concrete advice leaves little room for the researchers' own interpretations and because the following analysis of the advice (following subsections) is our main contribution---not the code system itself.

\begin{table}[tb]
    \centering
    \caption{Examples of document excerpts with concrete advice and assigned advice codes.}
    \label{tab:advice-paraphrases}
    \scriptsize
    \renewcommand{\arraystretch}{1.33}
    \setlength{\tabcolsep}{0.75\tabcolsep}
    \setlength{\defaultaddspace}{0.5\defaultaddspace} %
    \rowcolors{2}{white}{gray!10}
    \begin{tabularx}{\linewidth}{X}
        \toprule
        
        \textbf{Text in Document \& Assigned Advice Code}                             \\ \midrule
        \enquote{As a developer yourself, it’s your responsibility to provide users with the best and most secure experience possible.}                                                                                            	$\rightarrow$ Ensure authentication is secure and has good UX \\
        \enquote{Make sure to explain each \textins{2FA} step and keep in mind, that not everybody has the same technical knowlegde. For example, HubSpot added a description of Google Authenticator when the user would choose that option.} 
        $\rightarrow$ Explain each 2FA step                           \\
        \enquote{Consider Password-less Authentication \textelp{} Some applications are good candidates for a new approach: eliminating the password altogether. \textelp{} users \textelp{} forget their credentials, password-less authentication is an compelling option.}                                                                                                           $\rightarrow$ Use password alternatives                       \\
        \enquote{If your organisation wants to take its login game to the next level, you should definitely consider password alternatives.}                                                                                       $\rightarrow$ Use password alternatives                       \\
        \enquote{Send the user an email informing them that their password has been reset.}                                                                                                                                        $\rightarrow$ Notify users via email when password was reset  \\ 
        \bottomrule
    \end{tabularx}
\end{table}

As we focus on non-scientific online resources, we excluded \var{ScientificSourcesCount}~(\var{ScientificSourcesPercent}\%) documents being scientific papers; \var{VideoSourcesCount}~(\var{VideoSourcesPercent}\%) documents because they contained only a video or audio guide; another \var{UnrelatedSourcesCount}~(\var{UnrelatedSourcesPercent}\%) documents were not related to our research topic (\ie{} not about authentication and usable security); and two documents because they were not available anymore or a duplicate of an existing document. \var{NoAdviceSourcesCount}~documents contained no advice at all.
For the remaining \var{CorpusDocumentCountTotal-ScientificSourcesCount-VideoSourcesCount-UnavailableSourcesCount-NoAdviceSources-UnrelatedSourcesCount-DuplicateSourcesCount}~documents, we identified \var{DistinctAdviceCount}~distinct pieces of advice.

\subsubsection{Inclusion and Exclusion Criteria}
In the previous step, we extracted all advice for developers and obtained \var{DistinctAdviceCount}~paraphrased pieces of advice. 
We defined inclusion and exclusion criteria to focus only on advice relevant to this paper’s scope.
We excluded advice if any of the following criteria were met:
(1)~if the advice was not related to security, 
(2)~if the advice was not related to usability aspects and end-users, 
or (3)~if the advice did not cover authentication. 
The remaining advice was included as it is relevant for usable security of web authentication. 
Three researchers did this assessment independently and later resolved conflicts until they reached consensus. 

\paragraph{Advice Inclusion}
\var{DistinctAdviceCountRelevant}~distinct pieces of advice out of the initial \var{DistinctAdviceCount} covered usable security for web authentication and were therefore included for further analysis. 
For example, the advice to \enquote{Use a minimum of 8~characters} for passwords matched our inclusion criteria because passwords are a security and authentication feature, and their length affects usability~\cite{Komanduri:2011gl, Shay:2014:LongPasswords, Shay:2016:PasswordPolicies, Shay:2012:SystemAssignedPassphrases}. 
Overall, the \var{DistinctAdviceCountRelevant} pieces of advice matching the inclusion criteria occurred \var{TotalRelevantCodesInDocuments}~times across \var{UniqueDocumentsWithRelevantCodes}~different documents. 
Hence, only a small fraction of the online advice considers or affects usable security. 
Instead, most advice is only about security---not considering usability and human factors of security at all.

\paragraph{Advice Exclusion}
According to the above-mentioned criteria, we excluded \var{UnrelevantAdviceCount}~pieces of advice from further analysis because they did not cover usable security of web authentication, thus not being relevant for our research goals for one or multiple of the following reasons.
We excluded \var{UnrelevantNotSecurityAdviceCount}~distinct pieces that did not cover security, \eg{} \enquote{Provide live form validation} or \enquote{Only use single-column forms}. 
Similarly, we excluded \var{UnrelevantNotUsabilityAdviceCount}~distinct pieces of advice that were not relevant for end-users or usability. For example, \enquote{Use object-relational mapping~(ORM) to prevent SQL injections} or \enquote{Use bearer tokens,} are technical security details that users do not notice. 
We excluded another \var{UnrelevantNotAuthAdviceCount}~pieces of advice that covered usability and security but were not relevant to web authentication, \eg{} \enquote{Limit upload size for users}. 
We note that we focussed our analysis on advice in prose text and excluded any code snippets (occurring in 102~documents).

\begin{summaryBox}{Summary: Identifying \& Extracting Advice}
    We found and extracted \var{DistinctAdviceCount}~pieces of advice in \var{CorpusDocumentCountTotal}~documents. Focusing on web authentication advice with relation to usable security, we came up with \var{DistinctAdviceCountRelevant}~pieces of advice from \var{UniqueDocumentsWithRelevantCodes}~documents for further analysis.
\end{summaryBox}

\subsection{Advice Distribution \& Density}
\label{ssec:advice-distribution}

\begin{figure}[tb]
    \begin{subfigure}[c]{0.48\linewidth}
        \centering
        \includegraphics[width=\linewidth]{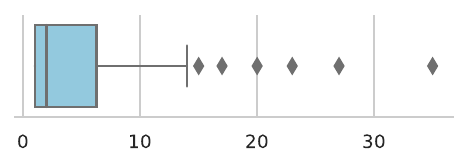}
        \caption{Number of pieces of advice per document.}
        \label{fig:boxplot-pieces-of-advice-per-document}
    \end{subfigure}
    \hfill
    \begin{subfigure}[c]{0.48\linewidth}
        \centering
        \includegraphics[width=\linewidth]{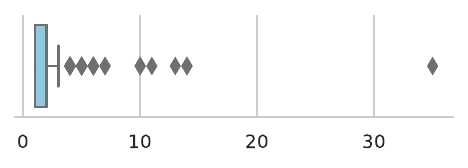}
        \caption{Occurrence frequency of advice (in number of documents).}
        \label{fig:boxplot-frequency-of-a-piece-of-advice-in-documents}
    \end{subfigure}
    \caption{Boxplots that describe the advice distribution.}
\end{figure}

As already stated, the advice was widely distributed across \var{UniqueDocumentsWithRelevantCodes}~documents. 
A document contained \var{advice.distribution.perDocument.mean}~different pieces of advice on average (median: \var{advice.distribution.perDocument.median}). However, the distribution was overall skewed (cf.\ \autoref{fig:boxplot-pieces-of-advice-per-document}): while the document with most advice contained \var{advice.distribution.perDocument.max}~different pieces of advice, the median was only \var{advice.distribution.perDocument.median}~different pieces of advice per document. That said, most documents provided only little advice, while a few were much denser. 
This was similar for the domains on which advice is provided. While individual domains (that contain advice at all) provided on average and median only \var{advice.distribution.perDomain.mean} and
\var{advice.distribution.perDomain.median}~different pieces of advice, respectively, the \emph{\gls{owasp} Cheat Sheets} contained most, having \var{advice.distribution.perDomain.max}~different advice imperatives (see \autoref{tab:top-domains-most-advice}). Considering that \emph{mobile-security.gitbook.io} by \gls{owasp} was also among the top domains, \gls{owasp} resources contained by far the most advice. 
However, authentication provider \emph{swoopnow.com}'s blog and the developer portals \emph{web.dev} and \emph{dzone.com} also contained a considerable amount of advice.

\begin{table}[tb]
\centering
\caption{Top-5 domains containing most distinct advice.}
\label{tab:top-domains-most-advice}
\footnotesize
\renewcommand{\arraystretch}{1.00}
\setlength{\tabcolsep}{0.75\tabcolsep}
\setlength{\defaultaddspace}{0.5\defaultaddspace} %
\rowcolors{2}{white}{gray!10}
\begin{tabular}{lr}
\toprule
                    \textbf{Domain} &  \textbf{\#Advice} \\
\midrule
cheatsheetseries.owasp.org &        55 \\
              swoopnow.com &        27 \\
mobile-security.gitbook.io &        27 \\
                   web.dev &        23 \\
                 dzone.com &        18 \\
\bottomrule
\end{tabular}
\end{table}

While most documents contained little advice, developers are not limited to considering only a single document. 
In line with that, our participants found \var{advice.distribution.perUpworker.mean}~different pieces of advice on average among multiple documents (median: \var{advice.distribution.perUpworker.median}). \var{advice.distribution.perUpworker.max}~different pieces of advice was the highest number of different advice imperatives found by a participant.
Looking into multiple information sources is also likely to yield new advice because \var{advice.distribution.frequency.onlyOneDocument} of \var{DistinctAdviceCountRelevant}~pieces of advice were only included in one document (cf.\ \autoref{fig:boxplot-frequency-of-a-piece-of-advice-in-documents}). 
On average, a piece of advice was only contained in \var{advice.distribution.frequency.mean}~documents (median: \var{advice.distribution.frequency.median}). 
However, a few pieces of advice were covered in a higher number of documents, as shown in \autoref{tab:top-advice}. 
\emph{Offer/Use 2FA/MFA} was the most frequent one, being contained in \var{advice.distribution.frequency.max}~documents. 

\begin{table}[tb]
\centering
\caption{Top-10 most frequent pieces of advice measured by occurrence in documents.}
\label{tab:top-advice}
\footnotesize
\renewcommand{\arraystretch}{1.00}
\setlength{\tabcolsep}{0.75\tabcolsep}
\setlength{\defaultaddspace}{0.5\defaultaddspace} %
\rowcolors{2}{white}{gray!10}
\begin{tabular}{lr}
\toprule
                                                                                                               \textbf{Advice} &  \textbf{Docs} \\
\midrule
                                                                                            Offer/Use 2FA/MFA. &     35 \\
                                                                              Enforce a password policy. &     14 \\
                                                                        Allow/Use 3rd-party SSO. &     14 \\
                                                   Use generic responses for auth.\  errors to not weaken security. &     13 \\
                                                      Require strong passwords. &     11 \\
                         Require re-authentication for sensitive actions (step-up auth). &     10 \\
     Reject most often used passwords and words from dictionaries. &     10 \\
Limit the number of login attempts. &      7 \\
                                    Protect against brute-force attacks by rate limiting. &      7 \\
                                                                                  Use password alternatives. &      6 \\
\bottomrule
\end{tabular}
\end{table}

All in all, advice is scattered, and most documents contained little advice, with a few exceptions. 
Hence, developers must consider multiple documents if they want to get a comprehensive set of advice. Looking at a single document would yield only a small subset of advice if not looking at one of the rare dense documents.

\begin{summaryBox}{Summary: Advice Distribution \& Density}
    While documents contained \var{advice.distribution.perDocument.mean}~pieces of advice on average, most only contained two or less. 
    However, some documents like \emph{\gls{owasp} Cheat Sheets} were denser.  \var{advice.distribution.frequency.onlyOneDocument} of \var{DistinctAdviceCountRelevant}~pieces of advice were only covered by a single document.
\end{summaryBox}

\subsection{Topics of Web Authentication Advice}
\label{ssec:categorization}

Early on, we noticed that the advice covered various topics. For further investigation, we took a closer look and identified \var{advice.categories}~different topics in our dataset's advice.

\subsubsection{Identifying Topics}
Three researchers independently used an inductive coding strategy to identify advice topics within the \var{DistinctAdviceCountRelevant}~pieces of advice and assign each advice paraphrase to its respective topic(s). 
Afterward, the researchers discussed each individual topic in a consensus session to create the final list of topics. 
It consists of \var{advice.categories}~different topics (cf.\ \autoref{tab:advice-categorization}). We assigned each topic a definition to achieve a reproducible assignment for each piece of advice and describe the topic. 
With the final list of topics, each coder independently coded all \var{DistinctAdviceCountRelevant}~pieces of advice. 
After the independent coding, the coders finally compared and resolved existing conflicts until they reached consensus. 
We allowed assigning multiple topics if a piece of advice fits several topics at the same time. 
For example, the advice to \enquote{Use password strength meters} covers \emph{passwords} and also implies \emph{\gls{ui}/\gls{ux}} requirements.

\subsubsection{Results}

Overall, we found \var{advice.categories}~different advice topics, as depicted in \autoref{tab:advice-categorization} (including each topic's definition, example advice, and prevalence counts). 
Within the \var{advice.categories}~topics, we found five topics corresponding to major authentication types: \emph{passwords}, \emph{\gls{2fa}/\gls{mfa}}, \emph{\gls{sso}/3rd-party login}, \emph{token-based authentication}, and \emph{biometrics}. 
Besides these five, some minor authentication types emerged that we grouped into an \emph{other} category (\eg{} TLS client certificates).
Advice on \emph{passwords} forms the most common authentication type, with \var{advice.category.passwords.unique}~distinct pieces of advice across \var{advice.category.passwords.documents}~documents.
The topic covers several aspects, like password policies, reset, entry, managers, or strength meters, and includes usability aspects, \eg{} toggling visibility in password fields. Advice on \gls{2fa} and \gls{mfa} is also popular (\var{advice.category.2fa/mfa.documents}~documents), followed by advice on \gls{sso}, third-party, and social logins. 
Token-based and biometric authentication were only rarely discussed.

The remaining six topics were more general and not about a specific authentication type. 
Usage recommendations, \ie{} what authentication mechanism to use, were very common (\var{advice.category.usage.documents}~documents). 
Another topic spanning all authentication types was session and account management (\eg{} account recovery). 
The remaining minor topics were about user and usability aspects. 
They covered general authentication \gls{ui} and \gls{ux}, how to interact with the user through notifications and messages, or how to educate users.

\newcommand\xxx{\par\hangindent1em\makebox[1em][l]{$\bullet$}}

\begin{table*}[tp]
    \caption{Categorization of advice in different topics.}
    \label{tab:advice-categorization}
    \centering
    \begin{threeparttable}
        \scriptsize
        \renewcommand{\arraystretch}{1.33}
        \setlength{\tabcolsep}{0.6\tabcolsep}
        \setlength{\defaultaddspace}{0.1\defaultaddspace} %
        \rowcolors{2}{white}{gray!10}
        \begin{tabularx}{\textwidth}{lllX>{\scriptsize\raggedright\itshape\arraybackslash}p{5.5cm}rrr}
            \toprule
            \multicolumn{3}{l}{\textbf{Advice Topic}} & \textbf{Definition} & \textbf{\textup{Advice Example}} & \textbf{\#\rlap{\tnote{1}}} & \textbf{\#Adv.\rlap{\tnote{2}}} & \textbf{\#Docs\rlap{\tnote{3}}} \\
            \midrule
            \multicolumn{3}{l}{\textbf{Authentication Type\tnote{\textdagger}}} &  &  &  &  &  \\ 
                {} & \multicolumn{2}{l}{\textbf{Passwords}} & Includes all advice, which directly mentioned a policy, a feature, or an implementation detail, that is directly or indirectly related to the creation, change, or processing of passwords. & \xxx Enforce a minimum password length of 8~characters. \xxx Reject passwords from breaches. & \var{advice.category.passwords} & \var{advice.category.passwords.unique} & \var{advice.category.passwords.documents} \\
                {} & \multicolumn{2}{l}{\textbf{\gls{2fa}/\gls{mfa}}} & Advice related to two-factor or in general multi-factor authentication and the different approaches but also backup and recovery and general considerations. & \xxx  Require adding a 2nd factor for features with higher security.
                    \xxx  Provide OTP as 2FA. & \var{advice.category.2fa/mfa} & \var{advice.category.2fa/mfa.unique} & \var{advice.category.2fa/mfa.documents}  \\
                {} & \multicolumn{2}{l}{\parbox[t]{2.4cm}{\textbf{\gls{sso}/3rd-Party \newline Login}}} & Recommendations related to the implementation or use of SSO, 3rd-party authentication services, or social logins, \eg{} via Facebook or Google. & \xxx Use OAuth 2.0 \xxx Stick to a few common social/third-party providers. &  \var{advice.category.sso} & \var{advice.category.sso.unique} & \var{advice.category.sso.documents} \\
                {} & \multicolumn{2}{l}{\parbox[t]{2.4cm}{\textbf{Token-Based \newline Authentication}}} & Authentication mechanisms based on tokens, their design, and usage. This also includes token formats such as Bearer or JWT. Cookie- and session-based are authentication also included. & \xxx Give tokens an expiration. \xxx Limit number of token usages. &  \var{advice.category.token} & \var{advice.category.token.unique} & \var{advice.category.token.documents}  \\
                {} & \multicolumn{2}{l}{\textbf{Biometrics}} & Includes all advice which directly is related to any kind of biometric authentication method. & \xxx Prefer fingerprint scanning above facial recognition. \xxx Use a traditional mechanism (e.g., password) beside biometrics. & \var{advice.category.bio} & \var{advice.category.bio.unique} & \var{advice.category.bio.documents}  \\
                {} & \multicolumn{2}{l}{\textbf{Other}} & Advice related to other authentication types, not belonging to one of the other authentication types. & \xxx Do not use TLS Client Authentication for public websites. \xxx Use passwordless authentication.  & \var{advice.category.other} & \var{advice.category.other.unique} & \var{advice.category.other.documents}  \\
            \midrule %
            \multicolumn{3}{l}{\parbox[t]{2.4cm}{\textbf{Usage \newline Recommendations}}} & Advice on how to choose or which authentication method to use. & \xxx Offer/Use 2FA/MFA. \xxx Use risk-based auth controls.  &  \var{advice.category.usage} & \var{advice.category.usage.unique} & \var{advice.category.usage.documents} \\ 
            \addlinespace 
            \multicolumn{3}{l}{\parbox[t]{2.4cm}{\textbf{Session/Account \newline Management}}} & Advice on management of sessions and accounts, including login and logout process, account creation/registration, recovery, blocking, user identifiers, etc. & \xxx Require re-authentication for sensitive actions (step-up auth). \xxx Invalidate all sessions after password change. & \var{advice.category.sessaccmngmnt} & \var{advice.category.sessaccmngmnt.unique} & \var{advice.category.sessaccmngmnt.documents}  \\ 
            \addlinespace 
            \multicolumn{3}{l}{\textbf{UI/UX}} & Concrete request to change, integrate or implement features and elements in the UI in a specific way, \eg{} how to design a password field. Also includes advice that directly mentions some aspect of user experience. & \xxx Use generic response code for auth/reg/recovery errors \xxx Don't prohibit pasting passwords. &  \var{advice.category.ui} & \var{advice.category.ui.unique} & \var{advice.category.ui.documents}  \\ 
            \addlinespace 
            \multicolumn{3}{l}{\parbox[t]{2.4cm}{\textbf{Notifications, Error-, \newline \& Warning Messages}}} & Advice on informing the user through notifications and messages in general, their design and usage, including push notifications, email, or SMS/text message, but also warnings, error messages. & \xxx Notify users via mail when password was reset. \xxx Avoid sending sensitive data to push notifications / backups / logs & \var{advice.category.notification} & \var{advice.category.notification.unique} & \var{advice.category.notification.documents}  \\ 
            \addlinespace 
            \multicolumn{3}{l}{\parbox[t]{2.4cm}{\textbf{User Education \newline \& Engagement}}} & Advice on engaging or educating users on different topics of authentication. & \xxx Educate users to create strong passwords. \xxx Encourage users to accept passwords from password managers. &  \var{advice.category.education} & \var{advice.category.education.unique} & \var{advice.category.education.documents} \\ 
            \addlinespace 
            \multicolumn{3}{l}{\textbf{Miscellaneous}} & Includes all advice that does not fit into any other category. & \xxx Use captchas after failed attempts. \xxx Involve stakeholders early on for security requirements. & \var{advice.category.misc} & \var{advice.category.misc.unique} & \var{advice.category.misc.documents}  \\ 
            \bottomrule
        \end{tabularx}
        \begin{tablenotes}
            \item [\textdagger] Container for sub-topics; containing no advice.
            \item [1] Advice occurrences in a topic.
            \item [2] Distinct pieces of advice in a topic.
            \item [3] Documents containing advice from a topic.
        \end{tablenotes}
    \end{threeparttable}%
\end{table*}

\begin{figure}
    \centering
    \includegraphics[width=0.9\columnwidth]{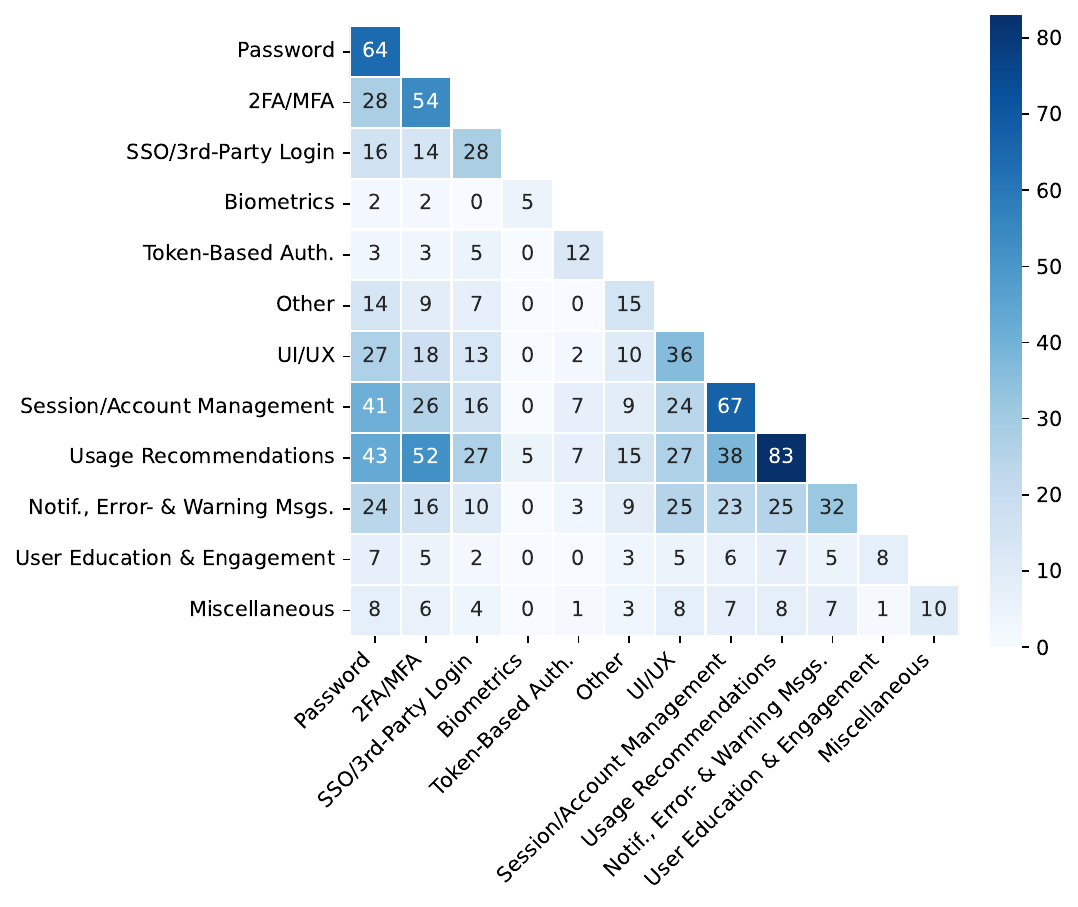}
    \caption{Co-occurrence of the topics from \autoref{tab:advice-categorization} within the same document.}
    \label{fig:cooccurrence-topics-in-documents}
\end{figure}

The co-occurrence heatmap of topics within a document in \autoref{fig:cooccurrence-topics-in-documents} reveals that authentication types are often not jointly covered within the same document. For example, while password advice is in \var{advice.category.passwords.documents}~documents, only in 28~documents both passwords and \gls{2fa}/\gls{mfa} are discussed together. This is similar for passwords and \gls{sso}/3rd-party login.
On average, a document touches \var{advice.distribution.topicPerDocument.mean}~different topics (median: \var{advice.distribution.topicPerDocument.median}). 
However, when just focusing on the authentication type topic subset, only \var{advice.distribution.topicAuthTypePerDocument.mean}~different authentication types are discussed. 
\var{advice.distribution.topicAuthTypePerDocument.onlyOneTopic}~documents cover only one authentication type.

\begin{summaryBox}{Summary: Advice Topics}
    Overall, we found a wide range of \var{advice.categories}~different advice topics. 
    Advice on passwords was by far the most frequent, followed by \gls{2fa}/\gls{mfa}.
    The topics are often discussed in isolation, \eg{} not discussing security benefits of \gls{2fa} when talking about passwords.
\end{summaryBox}

\subsection{Recommendable Advice}
\label{ssec:recommendable-advice}

Below, we provide details on advice we deem recommendable for developers. 
We consider authentication advice recommendable if it contributes to usable and secure authentication on the web.

\subsubsection{Identifying Recommendable Advice}
\label{sssec:recommendable-advice-identification}
We based our assessment on academic literature and official guidelines\extendedversion{ \cite{NIST:2017:SP-800-63B, lyastani2018better, sun2011, wiefling2019really, linden2005empirical, Abbott2020, reynolds_empirical_2020, european2015directive, Innocenti2021, hun2017, hardt2012oauth, okta-tokenlifetime2022, Naiakshina:2017, Simmons2021, golla2021driving, akhawe2013alice, harbach2013sorry, harbach2012measuringwarning, Gutfleisch:2022:usec-in-sdps, Cho2020, Lu2018, Huaman:2021:pwms, Adams:1999uj, Komanduri:2011gl, inglesant2010true, Habib2018-2, Bonneau:2012:sok, Schechter:2009:security-questions, Bonneau2015, Colnago2018, McDonald2021, Lee2021, Lee2020, Pearman:ccs17:ObservingPasswords, Stransky:2022:email, Apple:2017:FaceID, Whalen2022, Moradi2014, Yan2008, reCAPTCHA, w3c-url-for-changing, w3c-capchaalternatives}}{}.
In addition to references from related work, we searched for literature by checking the proceedings of top-tier venues, including IEEE S\&P, USENIX Security, ACM CCS, NDSS, USENIX SOUPS, ACM CHI, and official resources (\eg{} from \gls{nist}~\cite{NIST:2017:SP-800-63B}). 
This was complemented by searching on Google Scholar and a general online search. 
We also leveraged \emph{forward} and \emph{backward snowballing} to find further resources~\cite{Wohlin2014}, \ie{} resources citing or cited by an already obtained resource.

Based on these literature references, two authors assessed for each advice whether it could be considered recommendable. 
We used discussions to obtain consensus decisions. 
For transparency, we provide supporting references with their publication years to indicate their age.
Additionally, we provide written rationales that explain our reasoning within our replication package (\autoref{sec:replication-package}).

\extendedversion{%
We considered \var{advice.potentially-recommendable}~pieces of advice recommendable without having a reference or scientific evidence. 
Therefore, we excluded this advice in the results below and call it \emph{potentially} recommendable. 
Potentially recommendable advice includes advice such as \enquote{Use social login if building apps on top of Google or Facebook} or \enquote{Avoid login completely if not necessary}. 
For completeness, we provide the full list of potentially recommendable in the appendix, \autoref{tab:potentially-recommendable-advice}.
}{}

\subsubsection{Results}
We considered \var{advice.recommendable} of \var{DistinctAdviceCountRelevant}~pieces of advice (\var{advice.recommendable.perc}) recommendable. 
Overall, \var{advice.recommendable.affectedDocuments}~documents (\var{advice.recommendable.affectedDocuments.perc}) contained recommendable advice.
\autoref{tab:top-recommendable-advice} provides an overview of the most frequent recommendable advice. 
Comparing this to the overall most frequent advice (\autoref{tab:top-advice}) shows some minor differences. 
For example, we did not consider the advice to \enquote{Require strong passwords} recommendable because it does not provide concrete guidance.
However, we consider the related advice to \enquote{Enforce a password policy} recommendable since official institutions generally recommend password policies~\citewithyear{NIST:2017:SP-800-63B}, and there is scientific support for the use of password policies (while the concrete policy should be adapted to the context)~\citewithyear{Komanduri:2011gl, Lee2022}.

Among the advice that is only contained in one or two documents, we found highly recommendable advice for developers. 
For example, \enquote{Don't prohibit pasting passwords} (to not interfere with \glspl{pwm}~\citewithyear{Huaman:2021:pwms}) or to \enquote{Provide a password change feature.} 
The latter is crucial to change compromised passwords, but to our surprise only mentioned in two documents---maybe because it is too obvious.
Advice like \enquote{Don't force users to regularly change passwords} is recommendable, as it replaces the outdated advice to enforce regular password changes~\citewithyear{Habib2018-2, NIST:2017:SP-800-63B}.
Other advice also gives recommendations on specific pitfalls to avoid. For example, to not use TLS client certificates for public websites is recommendable because TLS client certificates are hard to use~\citewithyear{Parsovs2014}.

\begin{table}[tb]
\centering
\caption{Top-10 most frequent recommendable advice measured by occurrence in documents.}
\label{tab:top-recommendable-advice}
\scriptsize
\renewcommand{\arraystretch}{1.00}
\setlength{\tabcolsep}{0.75\tabcolsep}
\setlength{\defaultaddspace}{0.5\defaultaddspace} %
\rowcolors{2}{white}{gray!10}
\begin{tabular}{lr}
\toprule
                                                         \textbf{Advice} &  \textbf{Docs} \\
\midrule
                                             Offer/Use 2FA/MFA. &     35 \\
                                     Enforce a password policy. &     14 \\
                                       Allow/Use 3rd-party SSO. &     14 \\
Require re-authentication for sensitive actions (step-up auth). &     10 \\
  Reject most often used passwords and words from dictionaries. &     10 \\
                            Limit the number of login attempts. &      7 \\
          Protect against brute-force attacks by rate limiting. &      7 \\
                     Provide secure password recovery mechanism &      6 \\
                                Enforce minimum password length &      6 \\
                                  Use password strength meters. &      6 \\
\bottomrule
\end{tabular}
\end{table}

\begin{summaryBox}{Summary: Recommendable Advice}
    We found \var{advice.recommendable.perc} of the advice to be generally recommendable to developers, as it adheres to state-of-the-art research results or official guidelines. 
    Additionally, we consider \var{advice.potentially-recommendable.perc} of advice potentially recommendable that lacked references for evidence.
\end{summaryBox}

\subsection{Debatable Advice}
\label{ssec:debatable-advice}

Besides recommendable advice, we considered other advice to be debatable whether developers should follow it.
This includes if it contributes to insecure or unusable authentication on the web, is inconsistent with literature references, is only applicable in certain use cases, or different opinions on the advice exist.

\subsubsection{Identifying Debatable Advice}
We identified debatable advice as we did for recommendable advice (cf.\ \autoref{sssec:recommendable-advice-identification}).
Using the same references, two researchers assessed the advice and discussed debatable advice in a consensus session.
In cases without a supporting reference, we only tagged advice as debatable if a clear rationale existed. 
For traceability, we summarized the rationals  \extendedversion{of why advice is debatable in \autoref{tab:debatable-advice}}{in the extended version}. 
All \var{advice.questionable}~debatable pieces of advice are available in the replication package (\autoref{sec:replication-package}).

\subsubsection{Results}
We considered \var{advice.questionable} of \var{DistinctAdviceCountRelevant}~pieces of advice debatable. 
Overall, we found debatable advice in \var{advice.questionable.affectedDocuments}~documents (\var{advice.questionable.affectedDocuments.perc}).
\extendedversion{For a convenient overview, w}{W}e grouped similar debatable advice in \var{advice.questionable.groups}~groups (referred to as \emph{Dx}) and reported it along with the rationale why it is debatable and examples in \extendedversion{\autoref{tab:debatable-advice}}{the extended version}. 

Most debatable advice covered password policies~(D1) and was discussed in \var{advice.questionable.q1.documents}~documents. 
D1 is a good example of outdated advice. 
Outdated advice might have been considered correct in the past, but in the meantime, new insights emerged that render it incorrect. 
For example, we found advice to enforce complex composition rules, \eg{} passwords should consist of lower- and uppercase letters, numbers, and symbols. 
Another example is to \enquote{Force users to change passwords regularly.} 
However, research uncovered its negative effect on authentication usability and security~\citewithyear{Adams:1999uj, Komanduri:2011gl, inglesant2010true, Habib2018-2}. 
Likewise, official organizations such as \gls{nist} changed to less restrictive password policies in 2017 to improve usability and security compared to previous password policies~\citewithyear{NIST:2017:SP-800-63B}.
However, we also found up-to-date advice that is less strict and only suggests enforcing a minimal password length, as also recommended by \gls{owasp}~\citewithyear{OWASP:Authentication}. 

We found more outdated advice recommends the use of security questions~(D3), despite their usability~\citewithyear{Bonneau2015} and serious security shortcomings~\citewithyear{Bonneau:2012:sok, Schechter:2009:security-questions}, and official guidelines advising against them~\citewithyear{NIST:2017:SP-800-63B}.
We found other advice that harms (usable) security, such as not allowing to toggle masking in password fields and viewing the password in plain text~(D11), automatically terminating all sessions on password change without asking the user~(D6), or stating if a user entered an old but not their current password~(D7).
Some advice is more debatable, and its appropriateness depends on the use case. 
For example, consider the advice to use text messages (SMS) for authentication purposes~(D5). 
On the one hand, text messages have drawbacks like the possibility of losing access to a phone number~\citewithyear{McDonald2021, Lee2021}, SIM swapping attacks~\citewithyear{Lee2020}, and being interceptable~\citewithyear{NIST:2016:sms-discussion}. 
On the other hand, text messages are convenient as smartphones are ubiquitous. 
Similarly, PGP-encrypted emails for sending reset tokens~(D12) provide good security, but PGP has only poor adoption~\citewithyear{Stransky:2022:email}.
Some advice is challenging to assess, like fingerprints \emph{always} being preferable to facial recognition~(D13), while vendors like Apple changed from fingerprint to \emph{Face ID}~\citewithyear{Apple:2017:FaceID}. 

\begin{summaryBox}{Summary: Debatable Advice}
    In \var{advice.questionable.affectedDocuments.perc} of the documents, we found debatable advice that conflicts with research findings and official guidelines. 
    The debatable advice also contains instances of outdated advice, wrong advice, and advice that harms usable security.
\end{summaryBox}

\subsection{Contradicting Advice}
\label{ssec:contradicting-advice}

Besides recommendable (\autoref{ssec:recommendable-advice}) and debatable advice (\autoref{ssec:debatable-advice}), \ie{} advice in line or in conflict with literature references, we found several examples of contradicting advice within the document corpus itself.
To analyze those, we identified sets of advice that contradict each other (\eg{} ``Do X'' and ``Don't do X''). We refer to those as \emph{contradiction groups}. %
Overall, we identified \var{advice.contradicting.groups}~contradiction groups with a total of \var{advice.contradicting.unique}~distinct pieces of advice (cf.\ \autoref{tab:contradiction-groups}).
Most contradiction groups contained one pair of pieces of advice that directly contradict each other.
We illustrate our key findings below.

Most contradiction groups addressed passwords (C2, C4, C5, C7, C11, C13). For example, C4 is about whether to enforce regular password changes for users or not. That contradiction group's advice states two opposite approaches: to \enquote{Force users to change passwords regularly} (6~documents) and \enquote{Don't enforce regular password updates/changes for users} (2~documents). Similar disagreements can be found in password length maximum~(C2) and minimum~(C5) requirements or password visibility~(C7).
But contradicting advice also exists for other authentication areas, like session timeouts~(C3, C14), rate limiting~(C8, C12), \gls{2fa}~(C10), or general discussions on whether to use authentication at all~(C6).

\renewcommand\xxx{\par\hangindent1em\makebox[1em][l]{$\bullet$}\itshape\scriptsize}

\begin{table}[t]
    \caption{Overview of the \var{advice.contradicting.groups}~contradiction groups, each with a short description and advice examples (in italics).}
    \label{tab:contradiction-groups}
    \centering
    \begin{threeparttable}
        \scriptsize
        \renewcommand{\arraystretch}{1.15}
        \setlength{\tabcolsep}{0.4\tabcolsep}
        \setlength{\defaultaddspace}{0.0\defaultaddspace} %
        \rowcolors{2}{white}{gray!10}
        \begin{tabularx}{\linewidth}{>{\bfseries\arraybackslash}rXrr}
            \toprule
            \textbf{ID} & \textbf{Description \& Advice Examples} & \textbf{\#Adv.} & \textbf{\#Docs} \\
            \midrule
                C1         & \textbf{How authentication errors should be reported to the users.}         \xxx Use generic responses for authentication errors to not weaken security \xxx Tell if user types in an old password & \var{advice.contradicting.group.c13.advice} & \var{advice.contradicting.group.c13.documents} \\

                C2          & \textbf{Whether to set a maximum password length or not.}                   \xxx Set a maximum password length \xxx Don't set a maximum password length                                                                                                                           & \var{advice.contradicting.group.c2.advice} & \var{advice.contradicting.group.c2.documents} \\
                C3          & \textbf{Keeping users logged in or automatically log out.} \xxx Provide automatic logout/expire session \xxx Keep users signed in                                                                                                                                & \var{advice.contradicting.group.c7.advice} & \var{advice.contradicting.group.c7.documents} \\
                C4          &  \textbf{Whether to enforce regular password changes or not.}                 \xxx Force users to change passwords regularly \xxx Don't force users to regularly change/update passwords                                                                                        & \var{advice.contradicting.group.c1.advice} & \var{advice.contradicting.group.c1.documents} \\
                C5          & \textbf{ Different advice on minimum password length.}                       \xxx Use a minimum of 8 characters \xxx Use a minimum of 12--16 characters                                                                                                                             & \var{advice.contradicting.group.c3.advice} & \var{advice.contradicting.group.c3.documents} \\
                C6          & \textbf{Whether to use logins or not.}                                      \xxx Always require/enforce authentication \xxx Avoid login completely if possible/not necessary                                                                                                      & \var{advice.contradicting.group.c11.advice} & \var{advice.contradicting.group.c11.documents} \\
                C7          & \textbf{Whether to allow toggling password field visibility or not.}        \xxx Don't display passwords on the screen \xxx Allow users to toggle password visibility when typing it                                                                                              & \var{advice.contradicting.group.c4.advice} & \var{advice.contradicting.group.c4.documents} \\
                C8          & \textbf{Whether to use IP addresses or account for rate limits.}            \xxx Limit by account, not by IP \xxx Limit by IP                                                                                                                                                     & \var{advice.contradicting.group.c9.advice} & \var{advice.contradicting.group.c9.documents} \\
                C9         & \textbf{Using OAuth or OpenID Connect for authentication.}                  \xxx Implement 3rd-party authentication via OAuth\xxx Use OpenID Connect for authentication                                                                                                          & \var{advice.contradicting.group.c14.advice} & \var{advice.contradicting.group.c14.documents} \\
                C10         & \textbf{Which 2FA approach to choose.}                                      \xxx Provide OTPs as 2FA option \xxx Use mobile device push notifications as second factor                                                                                                            & \var{advice.contradicting.group.c12.advice} & \var{advice.contradicting.group.c12.documents} \\
                C11          & \textbf{Different speed requirements for password hashing.}                 \xxx Slow down hash functions \xxx Choose work factors so that hashing time \textless 1 sec                                                                                                           & \var{advice.contradicting.group.c6.advice} & \var{advice.contradicting.group.c6.documents} \\
                C12         & \textbf{Different rate limits to prevent brute force attacks.}              \xxx Limit to 3 attempts per account in a set timespan \xxx Limit to 20 guesses per minute and IP address                                                                                             & \var{advice.contradicting.group.c10.advice} & \var{advice.contradicting.group.c10.documents} \\
                C13          & \textbf{Entering passwords twice during registration or not.}     \xxx Let users confirm passwords by writing twice \xxx Don't ask for password confirmation                                                                                 & \var{advice.contradicting.group.c5.advice} & \var{advice.contradicting.group.c5.documents} \\
                C14          & \textbf{How long should tokens for authentication live.}                    \xxx Limit number of token usages \xxx Only allow one-time usage                                                                                                                                      & \var{advice.contradicting.group.c8.advice} & \var{advice.contradicting.group.c8.documents} \\
                C15         & \textbf{Whether to use CAPTCHAs or not.}                                    \xxx Use CAPTCHAs \xxx Use alternatives for CAPTCHAs & \var{advice.contradicting.group.c15.advice} & \var{advice.contradicting.group.c15.documents} \\
            \bottomrule
        \end{tabularx}
        \begin{tablenotes}
        \end{tablenotes}
    \end{threeparttable}%
\end{table}

Another important aspect is how contradicting advice occurs within a single document and within the document corpus, \ie{} across multiple documents.
Regarding \emph{intra-document} contradicting advice, we found this to be almost non-existent. Actually, only \var{advice.contradicting.conflicting_advice_in_same_document}~documents contain contradicting pieces of advice. 
However, this low intra-document contradiction rate is to be expected because it is very unlikely that an author is inconsistent within a document.
Analyzing the \emph{inter-document} advice contradictions, we identified \var{advice.contradicting.documents}~documents (\var{advice.contradicting.documents.percent}\%) that contained at least one piece of advice belonging to a contradiction group. 
That said, the document corpus contains many contradictions, indicating inconsistency and unreliability of online advice.

Contradiction groups contain opposite advice by definition.
Especially for debatable topics, contrary recommendations and opinions exist that are sometimes mixed with recommendable advice (\eg{} C1 or C4) and discussed in the document corpus. 
Therefore, the prevalence of debatable advice (\autoref{ssec:debatable-advice}) among the contradiction groups is considerably higher at \var{advice.questionable.contradicting.perc} compared to the overall \var{advice.questionable.perc}~rate of debatable advice.

\begin{summaryBox}{Summary: Contradicting Advice}
    We found \var{advice.contradicting.groups}~groups of contradicting advice.
    While it is almost non-existent within a single document, \var{advice.contradicting.documents.percent}\% of the documents contained some piece of contradicting advice. 
    Contradiction groups tend to contain more debatable advice.
\end{summaryBox}

\section{Discussion}
\label{sec:discussion}

Below, we summarize our key insights and discuss recommendations for developers, advice writers, and academia. 
This section concludes with a discussion of future work.

\subsection{Potential Impact of Advice on Authentication Deployments}  
Usable and secure web authentication is challenging, and previous work uncovered many issues, such as the widespread use of passwords in general~\cite{Bonneau:2012:sok, Bonneau2015a}, the deployment of complicated password policies~\cite{Lee2022, Komanduri:2011gl, Shay:2014:LongPasswords, Shay:2016:PasswordPolicies}, and the low adoption of password alternatives such as \gls{mfa} and FIDO2~\cite{Bonneau:2012:sok, Redmiles:2017:2FAMessages, AlQahtani2022, Colnago2018, Petsas2015, farke_you_2020, ghorbani_lyastani_is_2020}. 
Moreover, developers regularly depend on online resources to solve programming problems~\cite{Acar:2016ww, acar:2017:secdev, Acar:2017:InternetResources, Chen:2019:ICSE:HowReliable, Yang:2016:WhatSecurityQuestions} and might blindly trust information from blogs, documentation, or platforms such as \gls{so}~\cite{Fischer:2017, Fischer:2019:StackOverflowHelpful, Fischer:2021:ContentReranking}.

Our findings illustrate how online advice for web authentication might contribute to unusable and insecure authentication deployments. 
When developers seek online advice to solve challenges around online authentication, they likely stumble upon advice that is of limited help to make web authentication usable and secure. 
Most advice covered technical security implementation details that are not relevant for end-users, while only \var{DistinctAdviceCountRelevant} of \var{DistinctAdviceCount}~pieces of advice related to end-users' usable security (\autoref{ssec:extraction-filtering}).
Contradicting, debatable, and outdated advice might even have a negative effect.

For example, passwords were---despite their known potentially negative impact on security and usability---the most prevalent advice topic and often the only discussed authentication approach (\autoref{tab:advice-categorization}). 
The drawbacks and potential alternatives are rarely mentioned. 
We believe that such advice contributes to passwords' undiminished high popularity. 
Similarly, many websites do not follow best practices for password policies~\cite{Lee2022}, and advice is also often not in line with best practices (\autoref{ssec:debatable-advice}, \autoref{ssec:contradicting-advice}). 
Developers are not necessarily authentication or usable security experts. 
They might follow the advice unaware of the drawbacks and alternatives. 

Considering that developers rely on online resources for solving security programming challenges~\cite{Acar:2016ww, Fischer:2017}, we think that online advice is a major factor that affects web authentication usable security negatively, \eg{} outdated but still followed advice (\eg{} changing passwords regularly) or following advice that was always known to counteract usability and security (\eg{} using security questions).

\subsection{Challenges for Developers}
\label{ssec:challenges}
Based on our findings, we identified four main challenges for developers.
Overall, the results indicate a mixture of advice that could and should be followed and advice that should be carefully considered or avoided at all.  

\boldparagraph{Advice Diversity \& Priotrization.} The advice was highly diverse (\autoref{ssec:categorization}), \ie{} we found lots of advice on a single topic (\eg{} password policies). 
This and the high advice volume indicate an overload for developers and a lack of advice prioritization among its authors, as similarly found in related work on end-user advice~\cite{redmiles_comprehensive_2020}.

\boldparagraph{Scattered Advice.} The advice was also distributed over many documents (\autoref{ssec:advice-distribution}). 
While some documents contained more advice, many contained only a single piece of advice. 
Moreover, most advice was only covered in a single document. 
This makes it harder for developers to find relevant advice; they might have to consider many documents to find helpful advice for their needs.

\boldparagraph{Debatable Advice.} Debatable advice (\autoref{ssec:debatable-advice}) challenges developers, as they would have to recognize and not blindly follow detrimental advice. 
However, it cannot be assumed that developers 
(i)~are usable security experts and, for example, know which advice is outdated and should not be followed anymore; 
or (ii)~can recognize the consequences that poor usability of authentication features can have~\cite{Gutfleisch:2022:usec-in-sdps}.

\boldparagraph{Contradicting Advice.}
Finding contradicting advice (\autoref{ssec:contradicting-advice}) obliges the developer to decide which advice to follow and might be confusing. 
As with debatable advice, it is unclear whether developers have the skills to make an educated decision and assess related consequences.
Even when developers encounter only one piece of contradicting advice, chances are to blindly follow debatable or otherwise disadvantageous advice instead of best practices.

\subsection{Recommendations}

Given the potential impact of advice on authentication deployments and the associated challenges for developers, we advocate to rethink how advice is given and consumed to achieve authentication for end-users that is both more usable and secure. 
Based on our findings, we make the following recommendations for improving security advice for developers in the future.

\boldparagraph{Developers}
We found that online security advice on web authentication is diverse and distributed over many resources on the web. 
Overall, developers should consider their context and not follow advice blindly, as we found debatable advice that could not be unconditionally recommended (\autoref{ssec:debatable-advice}). %
Such assessments are mostly impractical for developers since they require lots of expertise, effort, and literature research.
Hence, reasonable resources that can be assumed to provide high-quality advice are necessary. 

We recommend developers look into the \gls{owasp} Cheat Sheets. 
According to our results, they contain the most advice and, therefore, are the most extensive resource, obviating the need for developers to look at many other resources that provide little or no advice. 
Another reason is that \gls{owasp} Cheat Sheets contain up-to-date advice specifically tailored to software practitioners.

\boldparagraph{Giving Advice}
We found several instances of outdated, incorrect, or otherwise debatable advice. 
Hence, authors should check their advice to avoid such problems before giving advice. 
This also implies regularly updating advice to prevent giving outdated advice. 
Unfortunately, checking advice is not straightforward, requiring expertise and considering resources such as the latest usable security research.
Since advice is highly dependent on specific contexts, advice givers should describe in which situations advice should be followed or not to prevent developers from following advice that does not suit their project.

As already discussed, we found parallels between advice and the authentication ecosystem. 
We argue that advice is related to and could influence the authentication landscape. 
Therefore, advice givers should carefully choose what authentication topics they write about. 
That does not mean that no advice on passwords should be given. 
However, the authors should focus on approaches that prevent weaknesses in passwords. 
For example, when writing about passwords, their drawbacks (\eg{} phishing, password reuse) should also be mentioned, and countermeasures (\eg{} \gls{2fa}) or alternatives (\eg{} FIDO2) should be outlined.

\boldparagraph{Official Institutions}
While we mentioned \gls{owasp} Cheat Sheets, the reader may wonder where official organizations stand as a source of advice. 
To our surprise, participants rarely reported official documents. 
For example, no \gls{nist} Special Publication was reported. 
This is especially remarkable as several documents referenced \gls{nist} guidelines and considered those authoritative. 
A closer look at \gls{owasp} Cheat Sheets~\cite{OWASP:CheatSheetSeries} and the related \gls{nist} SP~\cite{NIST:2017:SP-800-63B} quickly reveals major differences. 
For example, \gls{nist}'s PDF is much longer at 79~pages than \gls{owasp} Cheat Sheets. 
The language in \gls{nist}'s documents is also more abstract, \eg{} passwords are referred to as \enquote{memorized secret authenticators}~\cite{NIST:2017:SP-800-63B}.
We hypothesize that these factors lower the accessibility of \gls{nist} documents for developers and our participants did not report them for this reason. 

Nonetheless, official institutions like \gls{nist} are reputable, authoritative organizations whose documents contain reasonable advice. 
We propose that official institutions provide their advice in a form similar to \gls{owasp} cheat sheets. 
This could be done by condensing the most important advice from their extensive guidelines, standards, and other documents and presenting it in a developer-friendly and actionable way.

\boldparagraph{Authoritative Advice}
We obtained \var{DistinctAdviceCountRelevant}~distinct pieces of advice that showed high diversity and were scattered all over the internet. 
This imposes a challenge, as no developer would search \var{CorpusDocumentCountTotal}~documents for advice as we did. 
Hence, creating an authoritative source with helpful, actionable, carefully constructed web authentication advice for developers might improve accessibility, advice adoption, and end-user usable security. 
Moreover, this might prevent encountering debatable or contradicting advice.

It is yet unclear who should create and maintain authoritative advice. 
As already discussed, official institutions like \gls{nist} are considered authoritative but lack accessibility. 
Since this is mainly a presentation issue, we argue that official institutions could take this role. 
We found \gls{owasp} Cheat Sheets~\cite{OWASP:CheatSheetSeries} to be the most comprehensive knowledge base for authentication advice. 
Hence, \gls{owasp} should continue the Cheat Sheet Series project, while official institutions should start creating similar resources.
The advice extracted in this paper might be a starting point for creating (or extending) authoritative usable security authentication advice.

\boldparagraph{Knowledge Transfer: Academia to Practice}
Unsurprisingly, developers do not consider scientific resources like papers helpful (\autoref{fig:info-sources-helpfulness-ranking}). 
Developers are unlikely to read dozens of papers that might require expert knowledge or are hidden behind paywalls.
Instead, developers prefer online resources and heavily draw on contained advice~\cite{Acar:2016ww, acar:2017:secdev, Acar:2017:InternetResources}. 
However, academic papers yield new usable security knowledge and are the main communication channel for scientific results. 
Their impact is limited if they do not foster any change and gain real-world traction. 
For developer authentication advice, we can confirm the knowledge gap between academia and practice that \citeauthor{Gutfleisch:2022:usec-in-sdps} identified~\cite{Gutfleisch:2022:usec-in-sdps}.
To overcome this gap, researchers should 
aim to provide their knowledge in ways that developers can easily use and access. 
For example, new research results could be added to \gls{owasp} Cheat Sheets. 
While this can improve the adoption of advice that is grounded on research insights, the adoption is likely limited overall. 
Recent research indicates that Cheat Sheet advice is only partially followed~\cite{Innocenti2021}.

\subsection{Comparison with Related Work}
\label{ssec:discussion-comparison-rel-work}

While previous research investigated online security advice for both end-users~\cite{Hasegawa2022, redmiles_how_2016, fagan_why_2016, Ion2015, busse_replication_2019, Reeder2017, redmiles_comprehensive_2020} and software developers~\cite{Acar:2016ww, acar:2017:secdev, Acar:2017:InternetResources, Fischer:2017, Fischer:2019:StackOverflowHelpful, Fischer:2021:ContentReranking, Chen:2019:ICSE:HowReliable, Yang:2016:WhatSecurityQuestions, Barrera2022, Barrera2022a, Nguyen:2017, Geierhaas2022, Gorski:2018, reeder2011helping}, we are the first to explore developer advice on authentication that affects usable security for end-users.
As already discussed, we found similar themes compared to related work, like challenges around advice diversity and prioritization~\cite{redmiles_comprehensive_2020}, and a usable security knowledge gap between academia and industry~\cite{Gutfleisch:2022:usec-in-sdps}.
\citeauthor{redmiles_comprehensive_2020} investigated security advice regarding its quality (\eg{} actionability) in a user study~\cite{redmiles_comprehensive_2020}. 
As we performed a researcher evaluation, we refrained from evaluating aspects like developer actionability.
Such an assessment with developers, similar to \citeauthor{redmiles_comprehensive_2020}~\cite{redmiles_comprehensive_2020}, is potential future work and possible with our dataset. 
This might be interesting, as our overall impression is that quality is mixed, \eg{} some advice is highly concrete (\eg{} containing source code) while other is vague. 

Moreover, our work might provide some explanations for observations in related work. 
While not directly comparable, we see parallels to the findings of \citeauthor{Lee2022}~\cite{Lee2022}.
\citeauthor{Lee2022} found that only 13\% of websites followed all password policy best practices. 
A contributing factor to this can be developer advice---due to the plethora of different advice. 
Some advice might be recommendable, and others debatable, while potentially contradicting each other at the same time. 
Best practices identified by \citeauthor{Lee2022}, \eg{} block lists and strength meters, were also recommendations in our dataset but rarely occurred, reflecting the low adoption of best practices.
This low advice adoption might be caused by the challenges (\autoref{ssec:challenges}) that can become a burden for developers, ultimately resulting in a poor cost--benefit trade-off similar to the one outlined by \citeauthor{conf/nspw/Herley09} in 2009 for end-user security advice~\cite{conf/nspw/Herley09}.
Overall, we note that not only developers impact usable security of authentication.
Also, system administrators configuring authentication have an impact and draw on advice, \eg{} consulting standards and guidelines, as recently found by \citeauthor{Sahin2023}~\cite{Sahin2023}.
The authors identified similar challenges compared to our study, like administrators facing and following outdated recommendations.

In our dataset, the fraction of documents from Q\&A forums like \gls{so}  (\var{documents.type.forum/q&a.perc}) is lower than in related work (\eg{} 10--40\%~\cite{Acar:2016ww}).
As participants were free to choose any resource, we think the lower prevalence is due to different study characteristics. 
\gls{so} posts are likely most useful for programming as they provide code snippets for copy and paste~\cite{Fischer:2017}. 
However, our scenario is a more abstract, conceptual task  that does not include programming, while Acar et al.~\cite{Acar:2016ww} experimented with developers writing code.
Other factors might be self-reporting and social desirability biases that prevent participants from reporting \gls{so} because the platform is also known to contain low-quality or insecure answers~\cite{Fischer:2017, Chen:2019:ICSE:HowReliable}.
While we can only speculate about the reasons, future research is needed to investigate which resources software professionals prefer for abstract, conceptual tasks like ours.

\subsection{Outlook \& Future Work}
Changing advice and the authentication landscape can be considered long-term projects requiring suitable technical solutions. 
The endeavor to replace passwords has been an ongoing effort for decades~\cite{Adams:1999uj} that turned out to be hard~\cite{Bonneau:2012:sok} and has not been successful yet. 
Regarding advice, passwords are the prevailing topic. 
Therefore, we believe passwords to stay for more years to come. Users also still tend to prefer passwords~\cite{Zimmermann2020}.
However, the industry is trying to offer and shift to passwordless alternatives. 
In 2022, Apple, Microsoft, and Google announced plans to extend FIDO2 and introduce \emph{passkeys}~\cite{FidoAlliance:2022:Passkeys}. 
Passkeys aim to replace passwords by syncing FIDO credentials across the users' devices. 
While passkeys seem promising, future research is needed to study their effect. 
The advent of passkeys also calls for a future replication of this study to check how online advice evolves and how passkeys are covered.

While we qualitatively analyzed developer-reported web authentication advice in this paper, our work does not investigate developers' interaction with advice and advice impact on authentication usable security. 
Conceptional decisions (\eg{} whether \gls{2fa} is implemented or not) might not be decided by developers alone~\cite{Gutfleisch:2022:usec-in-sdps}. 
Hence, further investigation of advice adoption and decision processes is needed. 
One aspect of this is the prioritization of advice. 
We refrained from indicating recommendation strength for the advice, as a valid assessment requires a broader expert evaluation beyond the involved researchers. 
Future work could investigate recommendation strength with experienced software professionals.
We also propose to conduct a developer experiment focusing on the impact of advice on the final software to understand how advice should be given to developers. 
Similarly, future work could investigate the impact of code snippets, as we only focussed on textual (prose) advice.
We hope that our dataset can lay the foundation for future work in these directions.

\section{Conclusion}
\label{sec:conclusion}

In this work, we qualitatively analyzed \var{DistinctAdviceCountRelevant}~pieces of advice containing developer advice on usable and secure authentication.
We identify several major challenges for developers.
For example, the advice is scattered over many resources and, therefore, hard to find.
Despite their shortcoming, passwords are the most discussed authentication approach; the advice lacks the discussion of drawbacks and alternatives.
Moreover, unconsciously following advice can induce usable security problems, as we found instances of debatable, outdated, and contradicting advice.
Overall, our findings suggest that the advice 
potentially impacts authentication deployments.
Our recommendations can inform better advice giving to improve usable and secure authentication on the web.

\begin{acks}
    
We thank the anonymous reviewers for their valuable feedback and for helping us to improve this paper. 
We thank our study's participants for taking their time and allowing us to gain interesting insights. 
This research was funded by the \grantsponsor{dfg}{Deutsche Forschungsgemeinschaft (DFG, German Research Foundation)}{https://www.dfg.de/en/index.jsp} under \grantnum{dfg}{Germany's Excellence Strategy -- EXC 2092 \textsc{CaSa} -- 390781972}. %

\end{acks}

\printbibliography

\appendix 

\extendedversion{\begin{table}[tb]
\centering
\caption{Potentially recommendable advice.}
\label{tab:potentially-recommendable-advice}
\scriptsize
\renewcommand{\arraystretch}{1.00}
\setlength{\tabcolsep}{0.75\tabcolsep}
\setlength{\defaultaddspace}{0.5\defaultaddspace} %
\rowcolors{2}{white}{gray!10}
\begin{tabularx}{\linewidth}{Xr}
\toprule
                                                         \textbf{Advice} &  \textbf{Docs} \\
\midrule
  Use generic responses for auth.\ errors to not weaken security &     13 \\
                                      Require strong passwords &     11 \\
               Follow NIST password guidelines (\eg{} 800-63B) &      6 \\
              Avoid login completely if possible/not necessary &      3 \\
                             Make password change/reset simple &      3 \\
                                        Verify email address. &      2 \\
                       Never send the old password to the user &      2 \\
                            Limit session length intelligently &      2 \\
      Allow easy switching between "login" and "register" page &      2 \\
                Notify users via mail when password was reset. &      2 \\
       Use "login" and "sign up"; not "sign in" and "sign up". &      2 \\
    Take environmental factors into consideration (biometrics) &      2 \\
                Analyze which 2FA/MFA approach is best suited. &      2 \\
                                       Simplify authentication &      1 \\
                           Use long enough credentials/tokens. &      1 \\
                 Directly inform user why password is rejected &      1 \\
               Ensure authentication is secure and has good UX &      1 \\
                              Don't disclose account existence &      1 \\
   Use social login if building apps on top of Google/Facebook &      1 \\
   Provide an option to logout/destroy/invalidate the session. &      1 \\
                           Find a balance for lockout duration &      1 \\
                       Warn users when caps lock is activated. &      1 \\
     Put your login box or link clearly where users can see it &      1 \\
Encourage users to register multiple 2FAs for account recovery &      1 \\
 Don't lock account based on password reset/recovery requests. &      1 \\
                                  Provide backup codes for 2FA &      1 \\
             Issue multiple valid recovery codes to each user. &      1 \\
      Allow invalidating/revoking all existing recovery codes. &      1 \\
   Consider handling with parallel login from multiple devices &      1 \\
   Security measures must not prevent normal user from working &      1 \\
       Performance must stay adequate no matter how many users &      1 \\
         Verify user has active session when changing password &      1 \\
                                          Verify phone numbers &      1 \\
               Customize authentication workflow to your needs &      1 \\
               Chose only biometrics that fit your application &      1 \\
                               Enforce minimum/maximum lengths &      1 \\
                      Invalidate backup code after it was used &      1 \\
                  Make resetting passwords easy/obvious/simple &      1 \\
\bottomrule
\end{tabularx}
\end{table}
}{}
\extendedversion{\renewcommand\xxx{\par\hangindent1em\makebox[1em][l]{$\bullet$}}

\begin{table*}[tp]
    \caption{Overview of debatable advice, grouped in categories, each with representative advice examples.}
    \label{tab:debatable-advice}
    \centering
    \begin{threeparttable}
        \scriptsize
        \renewcommand{\arraystretch}{1.22}
        \setlength{\tabcolsep}{0.6\tabcolsep}
        \setlength{\defaultaddspace}{0.1\defaultaddspace} %
        \rowcolors{2}{white}{gray!10}
        \begin{tabularx}{\textwidth}{>{\bfseries\arraybackslash}r>{\bfseries\arraybackslash}L{1.8cm}X>{\scriptsize\raggedright\itshape\arraybackslash}p{4.1cm}rr}
            \toprule
            \textbf{ID} & \textbf{Title}                             & \textbf{Rationale}                                                                                                                                                                                                                                                                                                                                                                                                                                                                                                                                                                                                              & \textbf{\textup{Advice Example}} & \textbf{\#Adv.\rlap{\tnote{1}}} & \textbf{\#Docs\rlap{\tnote{2}}} \\ \midrule
            D1          & Password Policies                          & Much debatable advice is on password policies. Some recommends very strict password policies: character sets (like forcing at least a number, symbol, upper case letters, or any combination of those), high minimum length, or restricting maximum length. However, this is considered legacy according to latest guidelines (\eg{} by NIST~\citewithyear{NIST:2017:SP-800-63B}) and may also interfere with password manager usage~\citewithyear{Huaman:2021:pwms}. Similarly, regular password changes are still advised -- despite being unusable and known to have no significant security effect~\citewithyear{Adams:1999uj, Komanduri:2011gl, inglesant2010true, Habib2018-2}. Preventing password reuse is generally good advice, while hard to achieve across platforms and services. &     \xxx Use a minimum of 12--16 characters \xxx Generate random 16 digit passwords  \xxx Enforce alphanumeric characters \xxx Enforce punctuation in passwords \xxx Enforce a mixture of upper- and lowercase characters                  &  \var{advice.questionable.q1.advice} & \var{advice.questionable.q1.documents}                  \\
            D2          & Rate Limiting                              & Several pieces of advice are about limiting login attempts to prevent password guessing and credential stuffing~\citewithyear{NIST:2017:SP-800-63B}. However, its implementation can be debated and depends on the use case. We found advice to lock an account permanently or already after three attempts, which might lock legitimate users out of their accounts or could be leveraged for DoS attacks~\citewithyear{Lu2018}. Instead, temporary blocks might be more appropriate. However, one needs to choose an appropriate approach, \eg{} based on IP addresses, based on accounts, or regarding how many failed attempts are tolerated.     & \xxx Lock accounts after three or four failed login attempts \xxx Limit to three attempts per account in a set timespan \xxx Limit at 5 guesses/minute \xxx Limit by account, not by IP                    &     \var{advice.questionable.q2.advice} & \var{advice.questionable.q2.documents}                 \\
            D3          & Security Questions                         & Some advice (implicitly) suggests to use security questions. However, those pose a security threat as they are often easy to guess (\eg{} name of pet)~\citewithyear{Bonneau:2012:sok, Schechter:2009:security-questions}, have poor usability~\citewithyear{Bonneau2015} and official guidelines advise against their use~\citewithyear{NIST:2017:SP-800-63B}.                                                                                                                                                                                                                                                                                                                                                                                                                                    & \xxx Consider using a security question  \xxx Don't ask all security questions \xxx Don't use security questions alone for password reset                    &       \var{advice.questionable.q3.advice} & \var{advice.questionable.q3.documents}               \\
            D4         & Authentication Enforcement                 & Some advice is about enforcement, \eg{} always enforcing authentication or enforcing 2FA enrollment on registration. While this does not worsen security and might be an improvement, it is debatable if this is suitable and necessary in all situations. For example, many online shops work entirely with optional registration. Similarly, 2FA enforcement might only be reasonable for sensitive accounts but not important for mostly anonymous forum accounts, for example.                                                                                                                   & \xxx Enforce 2FA on registration \xxx Always require/enforce authentication                         &  \var{advice.questionable.q4.advice} & \var{advice.questionable.q4.documents}                    \\
            D5          & Usage of text messages (SMS)                               & Some advice is about using SMS for authentication, \eg{} sending 2FA codes. In general, SMS can be convenient for users, and they are therefore widely used~\citewithyear{Colnago2018}. However, there are some drawbacks like losing access to a phone number~\citewithyear{McDonald2021, Lee2021} or SIM swapping attacks~\citewithyear{Lee2020}. \gls{nist} also raises concerns~\citewithyear{NIST:2016:sms-discussion}. Overall, usage of SMS should be carefully thought about and alternatives considered.                                                                                                                                                                                                                                         & \xxx Send SMS instead of email \xxx Allow users to resend PIN code by SMS \xxx Notify users via text message/SMS \xxx Use SMS-based 2FA                      &     \var{advice.questionable.q5.advice} & \var{advice.questionable.q5.documents}                 \\
            D6          & Terminating Sessions after Password Change & To terminate all running sessions after password change or reset is understandable from a security perspective. Some advice recommends forcing this automatically. However, this could sign out authorized devices without any reason and results in tedious re-logins for the user. Instead, it might be more helpful to give the user a choice whether to invalidate active sessions or not.                                                                                                                                                                                                                                                                       & \xxx Invalidate all sessions after password change \xxx Force re-login after password reset                        &    \var{advice.questionable.q6.advice} & \var{advice.questionable.q6.documents}                  \\
            D7          & (Not) Entering Passwords Twice             & Whether passwords should be entered twice on registration and password changes or not is not well researched. Both variants have pros and cons, like a small probability of typos vs.\ tedious retyping of passwords. However, we believe that one-time-entry with an option to unmask the password field is more usable (if mistyped passwords can be reset/corrected).                                                                                                                                                                                                                                                                                              & \xxx Let users confirm passwords by writing it twice. \xxx Don't ask for password confirmation by writing it twice                         &   \var{advice.questionable.q7.advice} & \var{advice.questionable.q7.documents}                   \\
            D8         & Confusion: OAuth vs.\ OIDC                  & There is some confusion regarding when to use and what the OpenID Connect and OAuth protocols are. This is common on the internet, but OAuth refers to authorization, not authentication, as the official documentation states: \enquote{OAuth~2.0 is not an authentication protocol}~\citewithyear{OAuth:auth}. Therefore, the advice is factually wrong.                                                                                                                                                                                                                                        & \xxx Implement 3rd-party authentication via OAuth                          &  \var{advice.questionable.q8.advice} & \var{advice.questionable.q8.documents}                    \\
            D9          & Automatic Logout Time Span           & Some advice recommends short time spans to terminate a session automatically after inactivity (\eg{} for 2--3 minutes). While automatic logouts can be beneficial for security, too short time spans are a usability burden and might only be necessary for special applications (\eg{} banking). Official recommendations advise longer time spans, with a minimum of 15~minutes~\citewithyear{NIST:2017:SP-800-63B}.                                                                                                                                                                                                                                                        & \xxx Auto Logout: Mobile App 2--3 min \xxx Auto Logout: Web Apps <10 min \xxx Auto Logout: Secure workstation 10--15 min                         &   \var{advice.questionable.q9.advice} & \var{advice.questionable.q9.documents}                  \\
            
            D10          & Error Reporting for Login Failures         & Another recommendation is to display when someone entered an old password, when it was changed, or generally report that the entered password, username, or email is wrong. This is highly problematic, as it leaks information on account existence or credentials a user used in the past. As passwords are reused, this could weaken security of the user's other accounts~\citewithyear{Pearman:ccs17:ObservingPasswords}.                                                                                                                                                                                                                                                                                                                       & \xxx Notify users if they entered an old password \xxx If user typed an old password, output the date it was changed \xxx Output exactly which is wrong (username, email, or password)                         &     \var{advice.questionable.q10.advice} & \var{advice.questionable.q10.documents}                 \\
            D11         & (Not) Unmasking Passwords                        & There is still advice that indicates to always hide passwords in a password field. However, especially when users have long and unique passwords, the entry is error-prone~\citewithyear{NIST:2017:SP-800-63B}. Therefore, offering users an option to unmask the password improves usability. In some cases, displaying an unmasked password has no security implications (\eg{} private environment). To account for other situations, however, masking passwords should be the secure default.                                                                                                                                                                       & \xxx Don't display passwords on the screen                         &     \var{advice.questionable.q11.advice} & \var{advice.questionable.q11.documents}                 \\
            D12          & Using GPG Encrypted Mail                   & Using GPG encrypted mails, \eg{} to send users their reset tokens, seems to be a good idea in terms of security. However, the adoption of email encryption is poor~\citewithyear{Stransky:2022:email}. Therefore, this should rather be handled as an optional feature.                                                                                                                                                                                                                                                                                                                                                                                                                                                    & \xxx Send reset tokens in GPG encrypted mail                        &   \var{advice.questionable.q12.advice} & \var{advice.questionable.q12.documents}                   \\
            D13         & Fingerprint vs.\ Facial Recognition         & One advice is to prefer fingerprints to facial recognition. This advice is highly general, might depend on the technology in use, and could be wrong nowadays. For example, Apple changed from fingerprint \emph{Touch ID} to facial recognition \emph{Face ID} and evaluated the latter to be more secure~\citewithyear{Apple:2017:FaceID}.                                                                                                                                                                                                                                                                                                                                                                               &  \xxx Prefer fingerprint scanning to facial recognition                       &   \var{advice.questionable.q13.advice} & \var{advice.questionable.q13.documents}                   \\
            D14         & CAPTCHAs                                   & CAPTCHAs are a barrier that users have to overcome~\citewithyear{Whalen2022, Moradi2014, Yan2008}. Therefore, the advice to use CAPTCHAs should only be followed sparingly and if necessary~\citewithyear{NIST:2017:SP-800-63B}. Furthermore, this also depends on the CAPTCHA type, as some recent CAPTCHA technologies do not require any user interaction~\citewithyear{reCAPTCHA}.                                                                                                                                                                                                                                                                   & \xxx Use CAPTCHAs                         &   \var{advice.questionable.q14.advice} & \var{advice.questionable.q14.documents}                  \\
            \bottomrule
        \end{tabularx}
        \begin{tablenotes}
            \item [1] Number of distinct pieces of advice.
            \item [2] Number of distinct documents.
        \end{tablenotes}
    \end{threeparttable}%
\end{table*}
}{}
\section{Replication Package}
\label{sec:replication-package}

For transparency, potential replications of this study, and to support meta-research, we provide artifacts within a replication package.

\boldparagraph{Material}
We provide the following artifacts:
(1)~The scenario and task we used,
(2)~the materials used for recruiting, 
(3)~the consent form, 
(4)~the questionnaire from the developer survey, 
(5)~the list of advice that we extracted from the documents,
(6)~the detailed advice analysis results,
(7)~the set of documents, queries, and search results, 
and (8)~\extendedversion{this extended version of the paper}{the extended version of this paper}.
To ensure participant privacy, we do not provide any raw questionnaire responses.

\boldparagraph{Link}
The replication package is available at \url{https://doi.org/10.17605/OSF.IO/PFB5H}.

\extendedversion{\section{Scenario \& Task}
\label{sec:appendix-scenario-task}

We gave software developers the following scenario and task to search for online resources on web authentication: %

{\itshape %

For a research project on web development, we collect sources of information and advice that are used by software developers in software projects. As we focus on \emph{advice} sources that could be relevant for development, this task contains \emph{no actual programming}, but is about finding advice that might help you or others during development.

\noindent \textbf{Project Description}

Imagine you are working on a new software project called KeepYourHealth. KeepYourHealth is a new web-based health platform that patients can use to store and keep track of their health-related information. This includes diagnoses, medical images (\eg{} x-ray), and other data from doctors (similar to an electronic health record; but personal, voluntary, and under the user’s control) and can be supplemented with biometric and other data provided by the users. Patients can gather and upload biometric data such as heartbeat (data recorded by wearables, smartwatches, etc.) for early detection of cardiovascular diseases. KeepYourHealth should be compatible with both desktop and mobile browsers.

KeepYourHealth serves as a comprehensive and central instance to keep track of health, share selected data with doctors when needed, and help improve the health of its users. It will be used by people of any age, with different skills and should therefore be very \emph{easy-to-use}. At the same time, the platform needs to be \emph{secure}, as a recent survey found that potential users see the data (\eg{} medical data, biometric data, personal identifiable information) that is stored and processed with KeepYourHealth as highly sensitive.

\noindent \textsl{[Page break.]}

\noindent \textbf{Scenario}

You are responsible for the architecture and design of an easy-to-use and secure authentication workflow for the KeepYourHealth platform. For a kick-off meeting with other members of the development team, imagine that your job is to prepare a presentation with a \textbf{concept for the authentication system including user registration and login}. Therefore you have to do an online search.
 
\noindent \textbf{Requirements:}

\begin{itemize}
    \item The authentication process (and all other mechanisms) of KeepYourHealth has to be \textbf{secure}.

    \item The authentication process and its security features have to be \textbf{user-friendly} (provide a high level of usability for all kinds of end users like patients, doctors, other health care staff).

    \item You are free to make all necessary decisions including the choice of authentication methods, user interface, registration process, etc. As this project is in an early stage, you can choose any technology, programming languages, frameworks, and approaches that seem reasonable to you.
\end{itemize}

\noindent \textbf{Your Task:}

\begin{itemize}
    \item Prepare the kick-off meeting by \textbf{searching and collecting resources} from the web for \textbf{30~minutes} that you find \textbf{helpful and you might use for KeepYourHealth’s authentication process} as well as the requirements from above.
   
    \item Resources and advice can be all kinds of online documents (\eg{} blog posts, official documentation, guidelines, forum discussions, other websites).

    \item Report all resources and advice in the following table we provide, so your team can use these resources and its advice later in the project (like a “knowledge base”).

    \item Please include for each resource you found its URL and how you found it (\eg{} Google search query, a site you had bookmarked).
\end{itemize}

Please keep in mind that \textbf{usability} and \textbf{security} is in the \textbf{focus of this project}.

Here is an example of how the table looks like and how it could be filled: \textsl{[Example screenshot of the table.]}

\noindent \textsl{[Page break.]}

\begin{itemize}
    \item The \textbf{authentication process} for the KeepYourHealth platform should be \textbf{user-friendly} and \textbf{secure}.

    \item \textbf{Search 30 minutes for helpful resources on web authentication you might use} for the KeepYourHealth platform.

    \item It is sufficient if you \textbf{skim the resources roughly}. You don’t have to read those resources in detail.

    \item \textbf{Report} your \textbf{results} by filling the following \textbf{table}. You can also report resources where you found only a small part to be helpful for your task.

    \item You don’t have to prepare a presentation for the kick-off meeting. Currently, we are only interested in the resources you find online.
\end{itemize}

\begin{figure}[H]
    \centering
    \fbox{\includegraphics[width=0.97\linewidth]{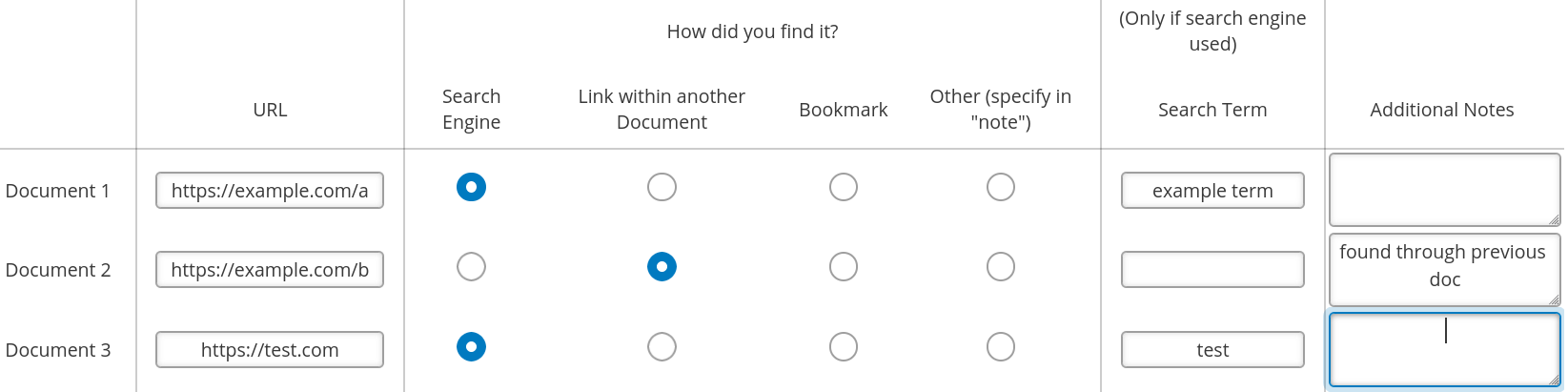}}
    \caption{A screenshot of the search task's table as used in the questionnaire.}
    \label{fig:screenshot-search-task-table}
\end{figure}

}}{}

\extendedversion{\clearpage

\section*{Supplementary material}

# ACM CCS'23 Artifact Appendix:
# "Make Them Change it Every Week!": A Qualitative Exploration of Online Developer Advice on Usable and Secure Authentication

Jan H. Klemmer 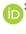*   Marco Gutfleisch 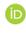†   Christian Stransky 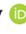C   Yasemin Acar 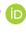‡

M. Angela Sasse 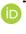†   Sascha Fahl 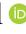C

*Leibniz University Hannover, Germany, {klemmer,stransky}@sec.uni-hannover.de
†Ruhr University Bochum, Germany, {marco.gutfleisch,martina.sasse}@rub.de
CCISPA Helmholtz Center for Information Security, Germany, sascha.fahl@cispa.de
‡Paderborn University, Germany, yasemin.acar@uni-paderborn.de

## 1 Artifact Appendix

### 1.1 Abstract

Our paper comprises human subjects research and a qualitative analysis of online advice for developers on usable and secure authentication. Therefore, we provide study materials and result data in our artifact (but no scripts with experiments, etc.). Hence, we apply *only* for the *Artifact Available* badge. Evaluation should be rather quick for the evaluators.

We provide the following artifacts: (1) The scenario and task we used, (2) the materials used for recruiting, (3) the consent form, (4) the questionnaire from the developer survey, (5) the list of advice that we extracted from the documents, (6) the detailed advice analysis results, (7) the set of documents, queries, and search results, and (8) the extended version of the paper. To ensure participant privacy, we do not provide any raw questionnaire responses.

Please see the contained README file for a detailed overview of the artifact's content.

### 1.2 Description & Requirements

This artifact has no specific requirements, as it only consists of some study materials and data, including all advice extracted in the paper.

#### 1.2.1 Security, privacy, and ethical concerns

None. There are no scripts to be run that. Hence, there is no risk for evaluators.

#### 1.2.2 How to access

The artifact is hosted at the Open Science Foundation (OSF) and is permanently available via the following DOI that is also linked in the paper: https://doi.org/10.17605/OSF.IO/PFB5H. (Or alternatively via direct OSF link: https://osf.io/pfb5h/.)

#### 1.2.3 Hardware dependencies

None.

#### 1.2.4 Software dependencies

None.

#### 1.2.5 Benchmarks

None.

### 1.3 Set-up

This artifact does not contain any scripts or experiments to be run. It just consists of the data and materials we provide. Therefore, the only set-up step is to download the artifact, so one could access and inspect it.

We note that one might want to use some program to conveniently view the CSV or Markdown files, although it is not necessary.

#### 1.3.1 Installation

The artifact can be accessed and inspected via the above-mentioned links. To download the artifact as a compressed ZIP archive from OSF, one can visit https://osf.io/

`pfb5h/files/osfstorage`, click "Download this folder," and extract the archive after downloading.

### 1.3.2 Basic Test

None.

## 1.4 Evaluation workflow

As we only apply for the *Artifact Available* badge, the evaluation should be straight-forward by just checking that the materials are available. This can be either done directly on the online platform or more convenient by downloading the materials locally. Overall, the evaluation should be quick, and we expect it to take 15–30 minutes at most.

## 1.5 Notes on Reusability

As the goal of this artifact is to provide the advice we extracted and analyzed in the paper, other researchers might want to inspect the advice or reuse some advice for their own research.

## 1.6 Version

Based on the LaTeX template for Artifact Evaluation V20231005. Submission, reviewing and badging methodology followed for the evaluation of this artifact can be found at `https://secartifacts.github.io/acmccs2023/`.

}{}

\end{document}